\documentclass[useAMS,usenatbib,usegraphicx,pdftex]{mn2e}
\usepackage{graphicx}
\usepackage{amsmath,amssymb}
\usepackage{pslatex}
\usepackage{float}
\usepackage{bm}
\bibpunct{(}{)}{;}{a}{}{,}

\newcommand{\xmm}{{\it XMM-Newton}}
\newcommand{\chandra}{{\it Chandra}}
\newcommand{\hst}{{\it HST}}

\title[Soft Extragalactic XRBs at the Eddington Threshold]{Soft Extragalactic X-Ray Binaries at the Eddington Threshold}
\author[Hannah M. Earnshaw, et al.]{\parbox{\textwidth}{Hannah M. Earnshaw$^{1}$\thanks{E-mail:
hannah.earnshaw@durham.ac.uk}, Timothy P. Roberts$^{1}$}\\
\\
\parbox{\textwidth}{
$^1$Centre for Extragalactic Astronomy, Department of Physics, Durham University, South Road, Durham, DH1 3LE, UK
}}

\begin{document}

\date{}

\pagerange{\pageref{firstpage}--\pageref{lastpage}} \pubyear{}

\maketitle

\label{firstpage}

\begin{abstract}

The luminosity range at and just below the $10^{39}$\,erg\,s$^{-1}$ cut-off for defining ultraluminous X-ray sources (ULXs) is a little-explored regime. It none-the-less hosts a large number of X-ray sources, and has great potential for improving our understanding of sources with $\sim$Eddington accretion rates. We select a sample of four sources in this Eddington Threshold regime with good data for further study; these objects possess a variety of soft spectral shapes. We perform X-ray spectral and timing analysis on the \xmm~and \chandra~data for these objects to gain insight into their accretion mechanisms, and also examine their optical counterparts using \hst~images. NGC~300~X-1 is a highly luminous and well-known example of the canonical steep power-law accretion state. M51~ULS exhibits a cool blackbody-like spectrum and is consistent with being an ultraluminous supersoft source (ULS), possibly a super-Eddington accreting object viewed at a high inclination through an optically thick outflowing wind. NGC~4395~ULX-1 and NGC~6946~ULX-1 have unusually steep power-law tails, for which we discuss a variety of possible physical mechanisms and links to similar features in Galactic microquasars, and we conclude that these sources are likely intermediate objects between the soft ultraluminous regime of ULXs and classic ULSs. 

\end{abstract}

\begin{keywords}
accretion, accretion discs -- black hole physics -- X-rays: binaries
\end{keywords}

\section{Introduction}
\label{sec:intro}

Our understanding of super-Eddington accretion has developed profoundly with the study of ultraluminous X-ray sources (ULXs), non-nuclear X-ray point sources with $L_{\rm X}>10^{39}$\,erg\,s$^{-1}$, the majority of which can be explained either as stellar-mass black holes (BHs) accreting at rates above the Eddington limit (e.g. \citealt{gladstone09}, see \citealt{feng11} for a review) or as highly super-Eddington neutron stars (e.g. \citealt{bachetti14, fuerst16, israel16}). A number of different X-ray spectral shapes and timing properties observed in ULXs can be unified with a model in which a clumpy, optically thick wind is driven off a geometrically thick, supercritical accretion disc, obscuring the hard central source at high accretion rates and/or high inclinations to the line of sight, and reprocessing its emission \citep{poutanen07, sutton13, middleton15a}. This model can also extend to ultraluminous supersoft sources (ULSs; \citealt{distefano04}), whose near complete lack of hard emission may be due to observing the source at relatively high inclinations or through an optically thick photosphere formed by the wind that completely envelops the source when the accretion rate is at its highest (e.g. \citealt{urquhart16, feng16}).

The highest luminosity ULXs are natural sources of interest as potential candidates for intermediate-mass BHs (IMBHs) and also simply because their luminosity allows for easier collection of sufficient data for in-depth analysis of nearer objects. However, there are a large number of sources in nearby galaxies at low ULX luminosities, or just below, that are less well-studied and a largely untapped resource for furthering our understanding of the lower-luminosity manifestations of super-Eddington accretion, as well as transitions from sub-Eddington to super-Eddington accretion regimes. We refer to this luminosity regime as the Eddington Threshold, encompassing objects with X-ray luminosity in the range $10^{38} < L_{\rm X} < 3\times10^{39}$\,erg\,s$^{-1}$. 

The main roadblock to the study of such objects is the relatively low amount of data we can collect using current missions, due to the generally low fluxes of these objects. Therefore for this study we select the best datasets available of this population at the Eddington Threshold as a proof of concept study for the science that could be achieved with next-generation missions such as {\it Athena}, whose larger collection area and higher sensitivity will make this class of objects far more accessible to investigation.

In Section~\ref{sec:sample} we detail how we selected our sample and summarise previous literature on these objects. We describe the data reduction process and analysis techniques in Section~\ref{sec:analysis}, and describe the results of X-ray spectral and timing analysis as well as optical counterpart photometry in Section~\ref{sec:results}. We discuss the implications of our results and draw comparisons between our sources and other known accretion regimes in Section~\ref{sec:disc}, and present our conclusions in Section~\ref{sec:conc}.

\section{Sample Selection}
\label{sec:sample}

\begin{table}
\caption{The properties of the sources and their host galaxies.} \label{tab:sources}
\begin{center}
\begin{tabular}{@{~}l@{~~}c@{~~~}c@{~~~}c@{~}c@{~}}
  \hline
  Name & R.A. \& Dec. & $d^a$ & $N_H^b$ & $E(B-V)^c$ \\
   & (J2000) & (Mpc) & ($\times 10^{20}$\,cm$^{-2}$) &  \\
  \hline
  NGC~300~X-1 & 00 55 10.0 $-37$ 42 12 & 1.83 & 4.04 & 0.0111 \\ 
  NGC~4395~ULX-1 & 12 26 01.5 $+33$ 31 31 & 3.98 & 1.86 & 0.0150 \\ 
  M51~ULS & 13 29 43.3 $+47$ 11 35 & 8.55 & 1.82 & 0.0309 \\ 
  NGC~6946~ULX-1 & 20 35 00.3 $+60$ 09 07 & 6.28 & 18.4 & 0.2944 \\ 
  \hline
\end{tabular}
\end{center}
$^a$The distance to the host galaxy, found by averaging the entries in the NED Redshift-Independent Distances database obtained from Cepheid standard candles for NGC~300 and NGC~4395, and from the tip of the red giant branch for M51 and NGC~6946. \\
$^b$The Galactic absorption column in the direction of this source, obtained by using HEASARC's $N_H$ tool (https://heasarc.gsfc.nasa.gov/cgi-bin/Tools/w3nh/w3nh.pl; \citealt{kalberla05}).\\
$^c$The Galactic extinction in the direction of the source, obtained by using the NASA/IPAC Infrared Science Archive's DUST tool (http://irsa.ipac.caltech.edu/applications/DUST/; \citealt{schlafly11}).
\end{table}

We selected our sample from a catalogue of ULXs and lower-luminosity X-ray point sources, created by matching the 3XMM-DR4 version of the \xmm~Serendipitous Source Catalogue \citep{rosen15} with the Third Reference Catalogue of Bright Galaxies (RC3; \citealt{devaucouleurs91}) using a method improving upon that in \citet{walton11}. Galaxy distances were refined by matching with \citet{tully88} or the NASA Extragalactic Database\footnote{http://ned.ipac.caltech.edu/} (NED) where possible, and the source luminosities were calculated using these distances and the error-weighted mean total flux from the \xmm~European Photon Imaging Camera (EPIC) instruments. This process resulted in a clean catalogue of 331 ULX candidates and 834 X-ray sources of lower luminosity associated with the same galaxies for the purpose of comparison (Earnshaw et al. in preparation). From this catalogue, we selected objects with peak luminosities at the Eddington Threshold ($10^{38} < L_{\rm X} < 3\times10^{39}$\,erg\,s$^{-1}$) according to the fluxes given in 3XMM-DR4, but with sufficient data for reasonably in-depth analysis. We defined this to be at least four \xmm~observations of the source, with multiple of these observations having counts in the thousands. We filtered this further to sources with variability of factor $\sim2$, to increase the chance of finding changes in accretion behaviour that could further increase the scientific value of the investigation.

These selection criteria leave us with four objects, all with soft spectra. We acknowledge this as an obvious selection bias, since extragalactic sources in the luminosity range we specify are only likely to have large numbers of counts available if they are dominated by soft emission. We summarise what is currently known about these sources below. A list of the sources and their host galaxy properties is provided in Table~\ref{tab:sources}.

\begin{table*}
\begin{minipage}{165mm}
\caption{The \xmm~and \chandra~observations for the sources in our sample.} \label{tab:observations}
\begin{center}
\begin{tabular}{cccccccc}
  \hline
  Name & ID$^a$ & Observation ID & Observatory & Instrument & Date & Exposure$^b$ & Off-axis Angle \\
   & & & & & & (ks) & (arcmin) \\
  \hline
  NGC~300~X-1 & X1 & 0112800101 & \xmm & MOS1/MOS2/pn & 2001-01-01 & 43.78/43.79/39.95 & 2.28 \\
   & X2 & 0112800201 & \xmm & MOS1/MOS2/pn & 2000-12-26 & 33.63/33.64/29.95 & 2.23 \\
   & X3 & 0305860301 & \xmm & MOS1/MOS2/pn & 2005-11-25 & 36.49/36.47/34.85 & 2.27 \\
   & X4 & 0305860401 & \xmm & MOS1/MOS2/pn & 2005-05-22 & 34.62/34.93/29.69 & 3.93 \\
   & X5 & 0656780401 & \xmm & MOS1/MOS2/pn & 2010-05-28 & 16.21/16.32/12.16 & 1.30 \\
   & C1 & 9883 & \chandra & ACIS-S & 2008-07-08 & 10.20 & 3.49 \\ 
   & C2 & 12238 & \chandra & ACIS-I & 2010-09-24 & 63.83 & 4.85 \\
   & C3 & 16028 & \chandra & ACIS-I & 2014-05-16 & 65.09 & 3.93 \\
   & C4 & 16029 & \chandra & ACIS-I & 2014-11-17 & 62.08 & 2.96 \\\\
  NGC~4395~ULX-1 & X1 & 0112521901 & \xmm & MOS1/MOS2/pn & 2002-05-31 & 15.08/15.10/10.08 & 3.07 \\
   & X2 & 0112522701 & \xmm & MOS1/MOS2/pn & 2003-01-03 & 8.18/8.19/6.57 & 2.26 \\ 
   & X3 & 0142830101 & \xmm & MOS1/MOS2/pn & 2003-11-30 & 103.65/104.06/98.16 & 1.89 \\
   & X4 & 0200340101 & \xmm & MOS1/MOS2/pn & 2004-06-02 & 75.88/76.69/68.43 & 13.24 \\
   & C1 & 5014 & \chandra & ACIS-I & 2004-08-07 & 33.14 & 12.4 \\\\
  M51~ULS & X1 & 0112840201 & \xmm & MOS1/MOS2/pn & 2003-01-15 & 20.66/20.67/19.05 & 1.65 \\
   & X2 & 0212480801 & \xmm & MOS1/MOS2/pn & 2005-07-01 & 35.04/35.77/24.94 & 1.18 \\
   & X3 & 0303420101 & \xmm & MOS1/MOS2/pn & 2006-05-20 & 39.60/39.66/30.90 & 0.66 \\ 
   & X4 & 0303420201 & \xmm & MOS1/MOS2/pn & 2006-05-24 & 29.77/29.76/23.18 & 2.82 \\ 
   & C1 & 3932 & \chandra & ACIS-S & 2003-08-07 & 48.61 & 2.52 \\
   & C2 & 13812 & \chandra & ACIS-S & 2012-09-12 & 159.54 & 3.32 \\
   & C3 & 13813 & \chandra & ACIS-S & 2012-09-09 & 181.57 & 3.32 \\
   & C4 & 13814 & \chandra & ACIS-S & 2012-09-20 & 192.36 & 3.56 \\
   & C5 & 13815 & \chandra & ACIS-S & 2012-09-23 & 68.07 & 3.64 \\\\
  NGC~6946~ULX-1 & X1 & 0200670101 & \xmm & MOS1/MOS2/pn & 2004-06-09 & 12.30/12.40/8.30 & 1.10 \\
   & X2 & 0500730101 & \xmm & MOS1/MOS2/pn & 2007-11-08 & 27.85/28.28/20.21 & 0.89 \\
   & X3 & 0500730201 & \xmm & MOS1/MOS2/pn & 2007-11-02 & 32.47/32.48/29.72 & 0.90 \\
   & X4 & 0691570101 & \xmm & MOS1/MOS2/pn & 2012-10-21 & 110.52/112.21/98.23 & 1.23 \\
   & C1 & 1043 & \chandra & ACIS-S & 2001-09-07 & 59.03 & 3.30 \\
   & C2 & 4404 & \chandra & ACIS-S & 2002-11-25 & 30.33 & 1.00 \\
   & C3 & 4631 & \chandra & ACIS-S & 2004-10-22 & 30.12 & 2.83 \\ 
   & C4 & 4632 & \chandra & ACIS-S & 2004-11-06 & 28.33 & 2.93 \\ 
   & C5 & 4633 & \chandra & ACIS-S & 2004-12-03 & 26.96 & 3.04 \\ 
  \hline
\end{tabular}
\end{center}
\vspace{-2mm}
$^a$A short source-specific observation ID, used for the remainder of this paper.\\
$^b$Sum of the good time intervals after removal of background flaring events. \xmm~values given as MOS1/MOS2/pn. 
\end{minipage}
\end{table*}

\begin{figure*}
\begin{center}
\includegraphics[width=4.3cm]{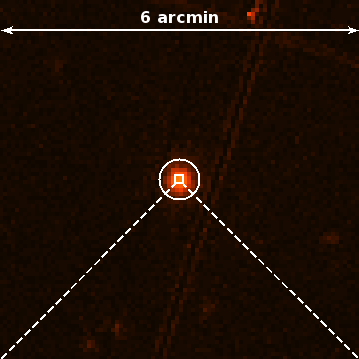}
\includegraphics[width=4.3cm]{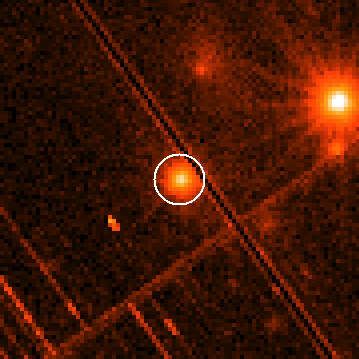}
\includegraphics[width=4.3cm]{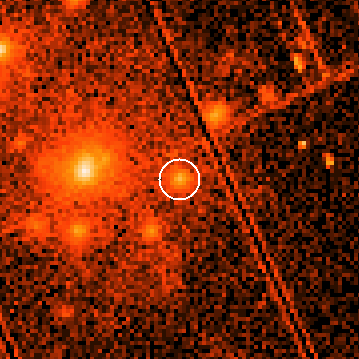}
\includegraphics[width=4.3cm]{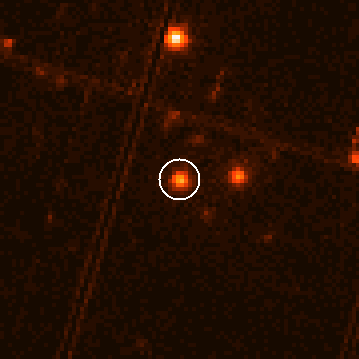}\vspace{0.02cm}
\includegraphics[width=4.3cm]{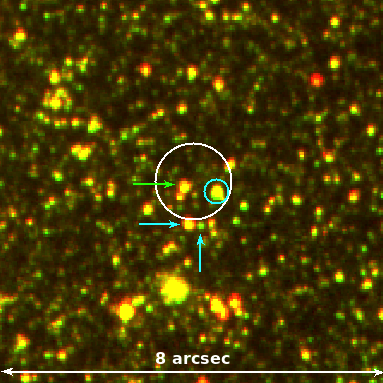}
\includegraphics[width=4.3cm]{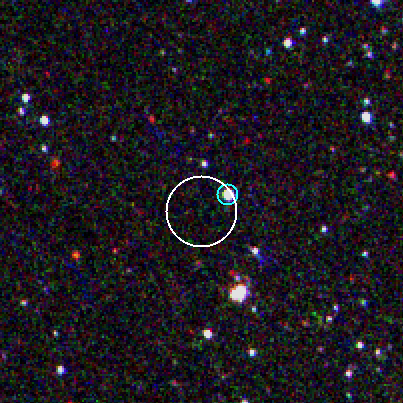}
\includegraphics[width=4.3cm]{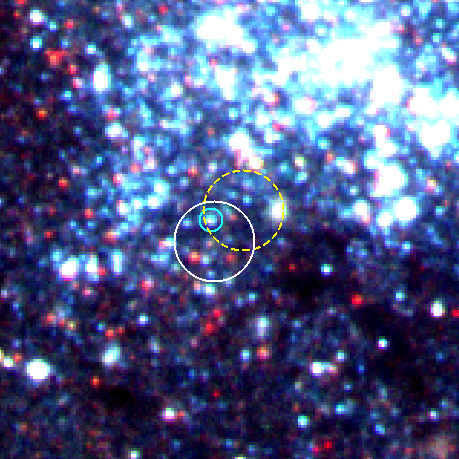}
\includegraphics[width=4.3cm]{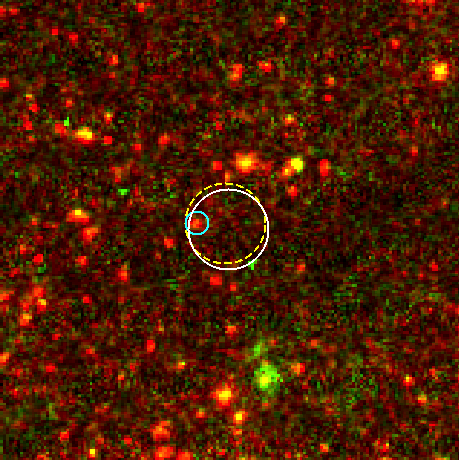}
\end{center}
\caption{The \xmm~EPIC-pn images (top) and astrometrically corrected \hst~images (bottom) for our sample of sources. The \xmm~boxes are 6$\times$6\,arcmin in size and the \hst~boxes are 8$\times$8\,arcsec, so fit well within the source positions in the top images. Source locations are as given in Table~\ref{tab:sources}. {\it Far left}, NGC~300~X-1 is marked with a 25\,arcsec white circle on \xmm~observation X1. The source position is marked with a 0.8\,arcsec white error circle on the \hst~image, with the {\textit F814W} band shown in red and the {\textit F606W} band shown in green. The WR candidate is marked with a cyan circle. We indicate the other two possible counterpart candidates identified by \citet{binder15} with cyan arrows, and one further bright star and possible candidate within our error circle with a green arrow. {\it Centre left}, NGC~4395~ULX-1 is marked with a 25\,arcsec white circle on \xmm~observation X3. The AGN in this galaxy can be seen towards the top right of the image. The source position is marked with a 0.7\,arcsec white error circle on the UV \hst~image, with the {\textit F438W} band shown in red, the {\textit F336W} band in green and the {\textit F275W} band in blue. The one counterpart we see, the same as identified in \citet{vinokurov16}, is marked with a cyan circle. {\it Centre right}, M51~ULS is marked with a 20\,arcsec white circle on \xmm~observation X2. The LLAGN can be seen on the left. The source position is marked with a 0.7\,arcsec white error circle and an alternate 0.7\,arcsec dashed yellow error circle on the \hst~image, with the {\textit F814W} band shown in red, the {\textit F555W} band in green and the {\textit F435W} band in blue. We indicate the counterpart identified in \citet{terashima06} with a cyan circle. {\it Far right}, NGC~6946~ULX-1 is marked with a 20\,arcsec white circle on \xmm~observation X3. The AGN in this galaxy can be seen to the right and X-1 \citep{pinto16} to the top. The source position is marked with a 0.7\,arcsec white error circle and an alternate 0.7\,arcsec dashed yellow error circle on the \hst~image, with the {\textit F814W} band shown in red and the {\textit F658N} band in green. A single possible counterpart is marked with a cyan circle.}
\label{fig:images}
\end{figure*}

\subsection{NGC 300 X-1}
\label{sec:ngc300}

Of our sample, 2XMM J005510.0-374212 (henceforth NGC~300~X-1, from \citealt{carpano07}) is by some margin the best-studied object. It is a well-known Wolf-Rayet/BH binary within the eastern spiral arm of its host galaxy. The radial velocity obtained from optical spectroscopic observations has been used to place limits on the mass function of the system, and derive a BH mass of $20\pm4$\,M$_{\odot}$ \citep{crowther10}. Its high mass and low distance, coupled with the fact that it has most frequently been observed in the very luminous steep power-law state (e.g. \citealt{carpano07}, \citealt{binder11}), makes it unsurprising that it has the luminosity and data quality sufficient to appear in our sample. 

NGC~300~X-1 has been observed five times with \xmm, for a total observing time of approximately 170\,ks. It has also been observed five times with \chandra~for a total of over 220\,ks and, recently, with the Hubble Space Telescope ({\it HST}) Advanced Camera for Surveys (ACS) in the {\textit F606W} and {\textit F814W} bands. While it has previously been assumed that the Wolf-Rayet star is the binary companion, the \hst~images also show a number of other possible companion stars, including an AGB star candidate and a high-mass main sequence star \citep{binder15}.

\subsection{NGC 4395 ULX-1}
\label{sec:ngc4395}

While the least variable of our sample between \xmm~observations, 2XMM J122601.4+333131 (henceforth NGC~4395~ULX-1, from \citealt{liu05}) has a large amount of \xmm~data available, with four \xmm~observations not dominated by background flaring, each with source counts in the multiples of thousands and one in excess of 30,000, with a total observing time of $\sim214$\,ks. It is also detected in three \chandra~observations for a total of $\sim44$\,ks. It has previously been identified as a ULX in the catalogues of \citet{roberts00}, \citet{liu05} and \citet{swartz11}, and found to have a soft power-law-like spectrum by several studies, albeit one with significant residuals, possibly from hot diffuse gas \citep{feng05, stobbart06}. It shows significant long-term variability according to {\it Swift} data \citep{kaaret09}, and was recently found to exhibit a period of 62.8 days \citep{vinokurov16}.

The south-east portion of the galaxy containing NGC~4395~ULX-1 has been previously imaged by \hst~in six bands with the WFC3 and ACS instruments, although for the optical bands {\textit F814W}, {\textit F555W} and {\textit F435W}, the source lies within the Advanced Camera for Surveys' chip gap. NGC~4395~ULX-1 has a singular optical counterpart with a blue power-law spectral energy distribution over the {\textit F275W}, {\textit F336W} and {\textit F438W} bands. Its optical spectrum shows evidence of broad He~{\sc ii} emission, consistent with other ULX counterparts \citep{vinokurov16}.

\subsection{M51 ULS}
\label{sec:m51}

2XMM J132943.2+471134 (henceforth M51~ULS, also called M51~ULX-2 in \citealt{terashima04} but not to be confused with M51~ULX-2 in \citealt{urquhart16b}) is located in the western spiral arm of M51, a very well-studied galaxy with a large population of ULXs and an abundance of multi-wavelength data. It has been observed six times with \xmm~although M51~ULS is only detected in four of them, totalling $\sim154$\,ks of observing time. It has also been the subject of a deep \chandra~observing campaign, placing the total observing time at just under 750\,ks. Previous studies analysing the ULX population of M51 have discovered this source to be very soft \citep{distefano04}, able to be fit with a cool accretion disc model ($\sim0.1$\,keV; \citealt{terashima04}) with a small amount of additional hard emission that can be fit with a combination of a power-law tail and a MEKAL thermal plasma component \citep{dewangan05}.

A deep \hst~observation as part of the Hubble Heritage Project in 2005 allowed the identification of a probable optical counterpart to M51~ULX-2, found to be consistent with being an OB-supergiant type star, making M51~ULS a high-mass X-ray binary \citep{terashima06}. 

\subsection{NGC 6946 ULX-1}
\label{sec:ngc6946}

Finally, 2XMM J203500.1+600908 (henceforth NGC~6946~ULX-1, from \citealt{liu05}, but not to be confused with NGC~6946~X-1 in \citealt{pinto16}) is a soft source in an inner eastern spiral arm of its host galaxy, and the only object of our sample with peak luminosity in the traditional ULX range ($L_{\rm X} > 10^{39}$\,erg\,s$^{-1}$) according to the 3XMM-DR4 data. While the galaxy has been observed by \xmm~11 times, the source is only detected with sufficient data for spectral analysis in four of them, although those four observations total over 185\,ks of observing time. Additionally, it has been observed for $\sim210$\,ks with \chandra~over six observations. Like NGC~4395~ULX-1, it has previously been found to be a soft source, by \citet{devi08} who claim it to be an IMBH due to being acceptably fitted with a cool accretion disc.

NGC~6946~ULX-1 has been observed in two \hst~bands with the ACS instrument, {\textit F814W} and {\textit F658N}. 

\section{Data Reduction and Analysis}
\label{sec:analysis}

We conducted our analysis on the objects listed in Section~\ref{sec:sample} using archival \xmm~and \chandra~observations. A list of the observations used in this investigation -- those where the source is detected with enough data for spectral analysis, approximately 400 counts at minimum -- is given in Table~\ref{tab:observations}.

We reduced the \xmm~data using {\sc v13.5.0} of the \xmm~Science Analysis System (SAS) software package. We first created calibrated and clean event lists by running the tasks {\sc emproc} and {\sc epproc}, then removed intervals dominated by background flaring, defined as a $>10$\,keV count rate greater than 0.35\,ct\,s$^{-1}$ for the EPIC-MOS data and a 10--12\,keV count rate greater than 0.4\,ct\,s$^{-1}$ for the EPIC-pn data. Using {\sc evselect}, we selected source events for spectra and light curves with {\sc pattern $<=$ 12} for the MOS cameras and {\sc pattern $<=$ 4} and {\sc flag == 0} for the pn camera from a circular region around the source with radius 20\,arcsec for M51~ULS and NGC~6946~ULX-1, and 25\,arcsec for NGC~300~X-1 and NGC~4395~ULX-1. Background counts were taken from equally-sized regions outside the host galaxy and on the same chip at a similar distance from the readout node. Redistribution matrices and auxiliary response files were generated using the {\sc rmfgen} and {\sc arfgen} tasks respectively. The spectral data was grouped into at least 20 counts per bin to allow for Gaussian statistics, and oversampling limited to a maximum of three groups per spectral resolution FWHM by setting {\sc oversample=3} in the {\sc specgroup} task. 

The \chandra~data was reduced using {\sc v4.7} of the Chandra Interactive Analysis of Observations (CIAO) software package and reprocessed using {\sc chandra\_repro} to produce up-to-date event lists. Spectra and light curves were extracted using the {\sc specextract} and {\sc dmextract} routines respectively, with the same binning as the \xmm~data, using the settings {\sc weight=no} and {\sc correctpsf=yes} given that each of our objects is a point source. We extracted source counts from 3\,arcsec radius circular regions around the source, except for one instance of an extended PSF when the source was off-axis, when we used a source region of 5\,arcsec. Background counts were extracted from an annulus around the source with inner radius equal to the radius of the source region and outer radius of 20\,arcsec. 

As well as exhibiting significant long-term variability, each of our sources except for NGC~4395~ULX-1 have at least one \xmm~observation flagged as variable in 3XMM-DR4  (having a $\chi^2$ probability of constant rate $<1\times10^5$). We examined the timing properties of the sources by creating power spectra. We checked that the power spectrum was stationary across all observations, then we divided all of the available lightcurves for a source from a single telescope into approximately 20 equal-length segments. Then we averaged over the periodograms of each segment to obtain a power spectrum for each of \xmm~and \chandra, which we normalised into units of squared fractional rms per frequency interval. In the case of NGC~300~X-1, for which there is sufficient signal above the white noise level, we repeated this procedure for five different energy bands and integrated the resultant power spectra to create an rms spectrum.

\begin{figure*}
\begin{center}
\vspace{-0.5cm}
\includegraphics[width=9cm]{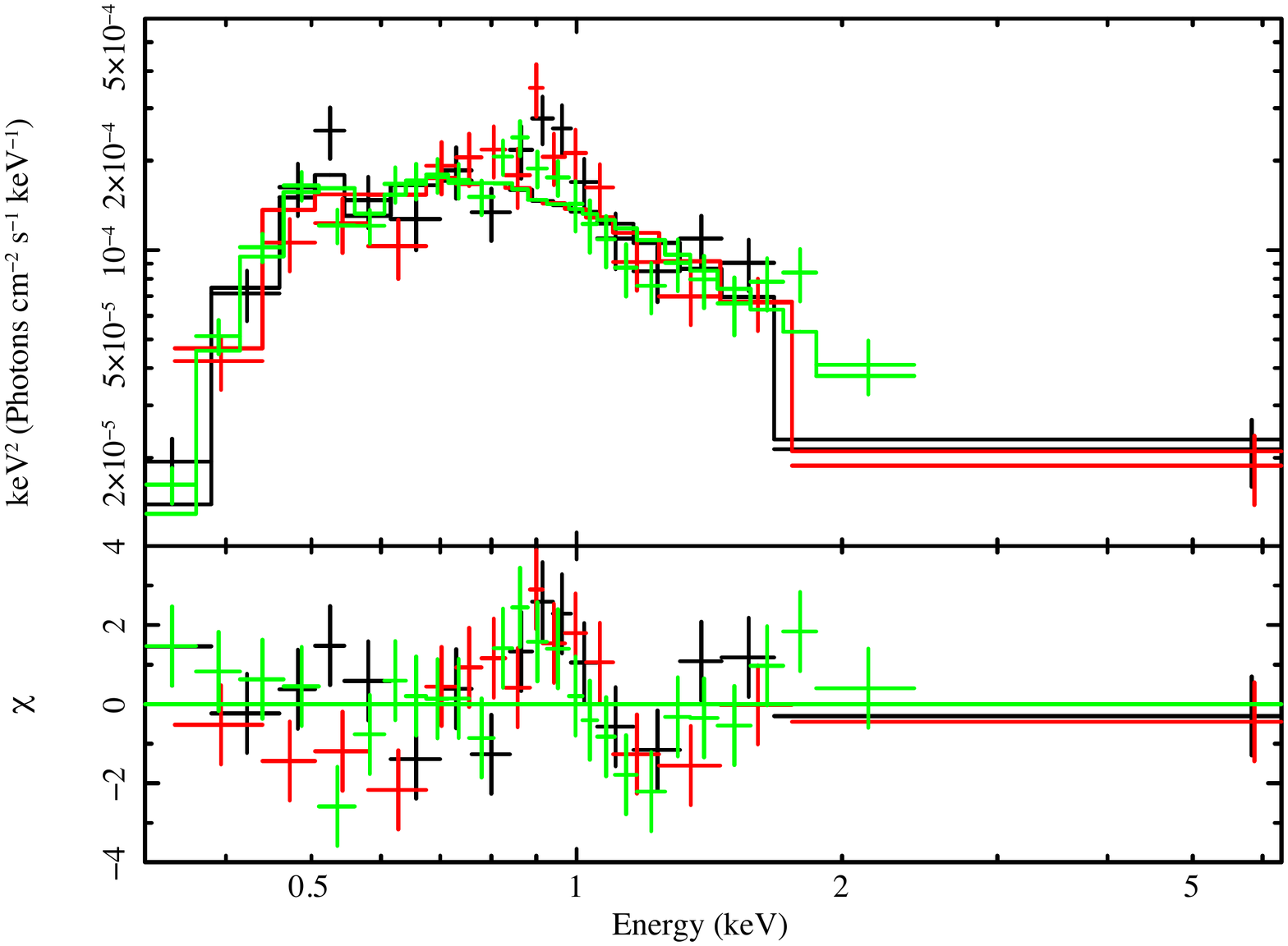}\hspace{-0.5cm}
\includegraphics[width=9cm]{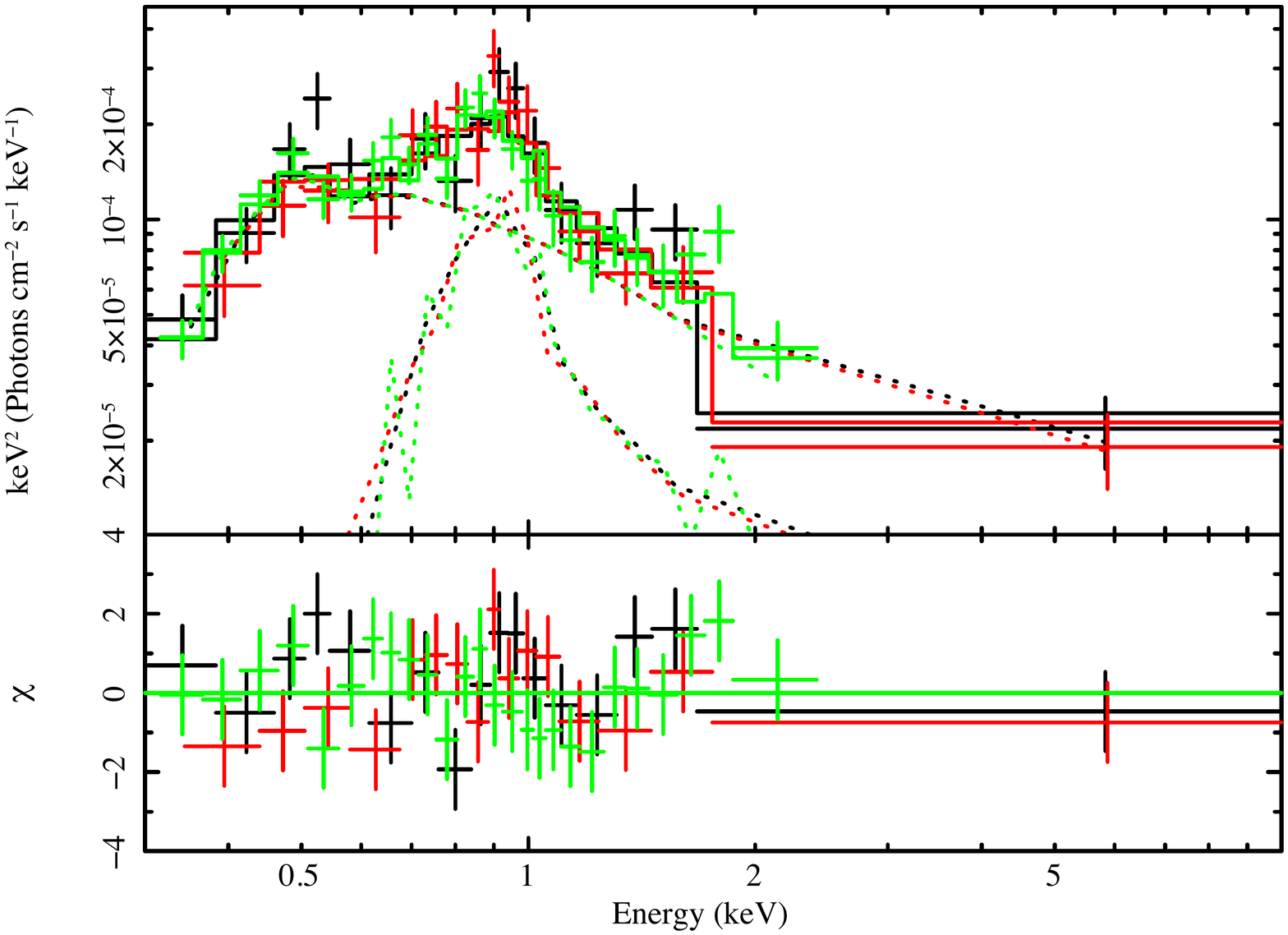}\vspace{-1cm}
\includegraphics[width=9cm]{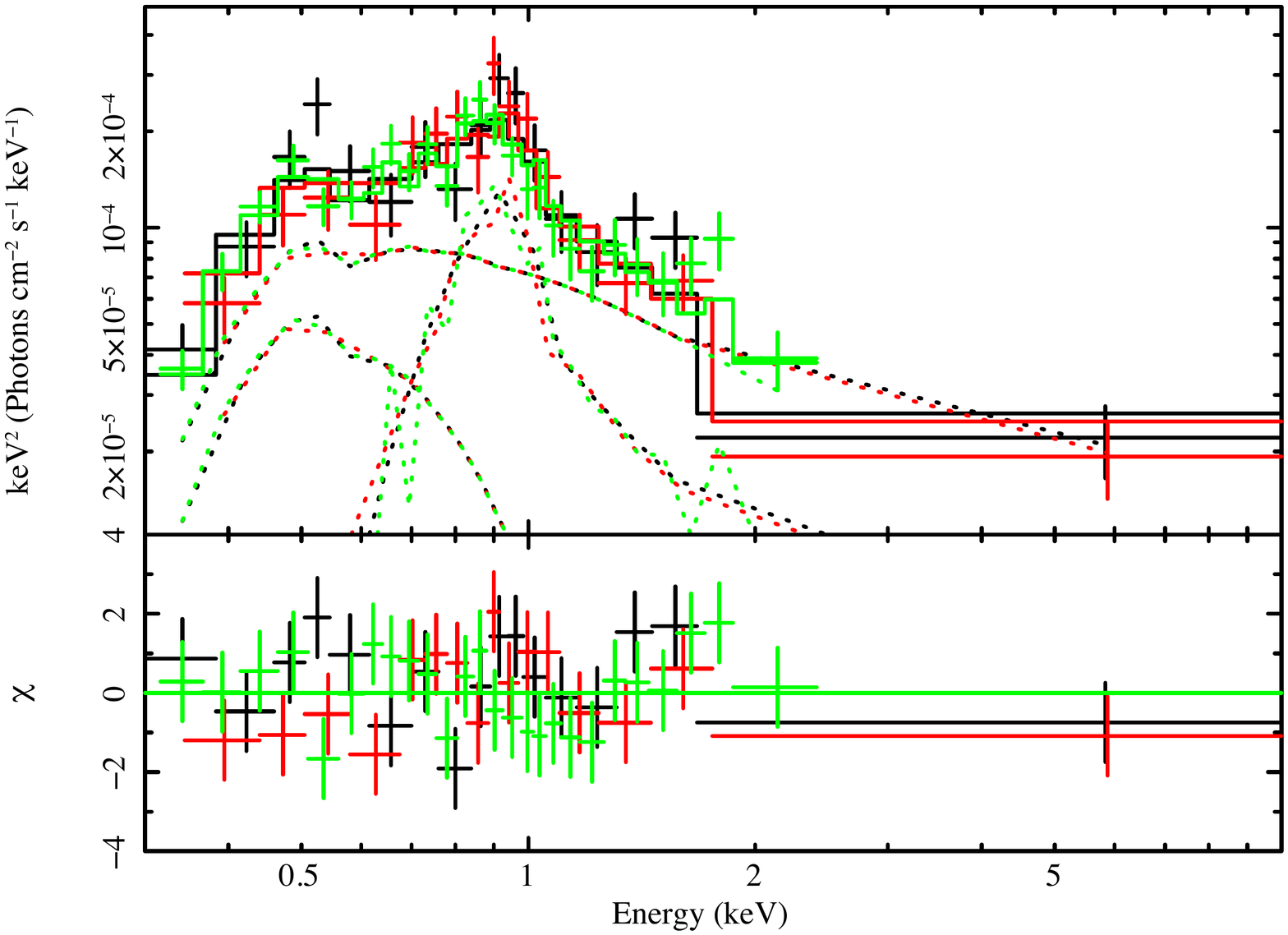}\hspace{-0.5cm}
\includegraphics[width=9cm]{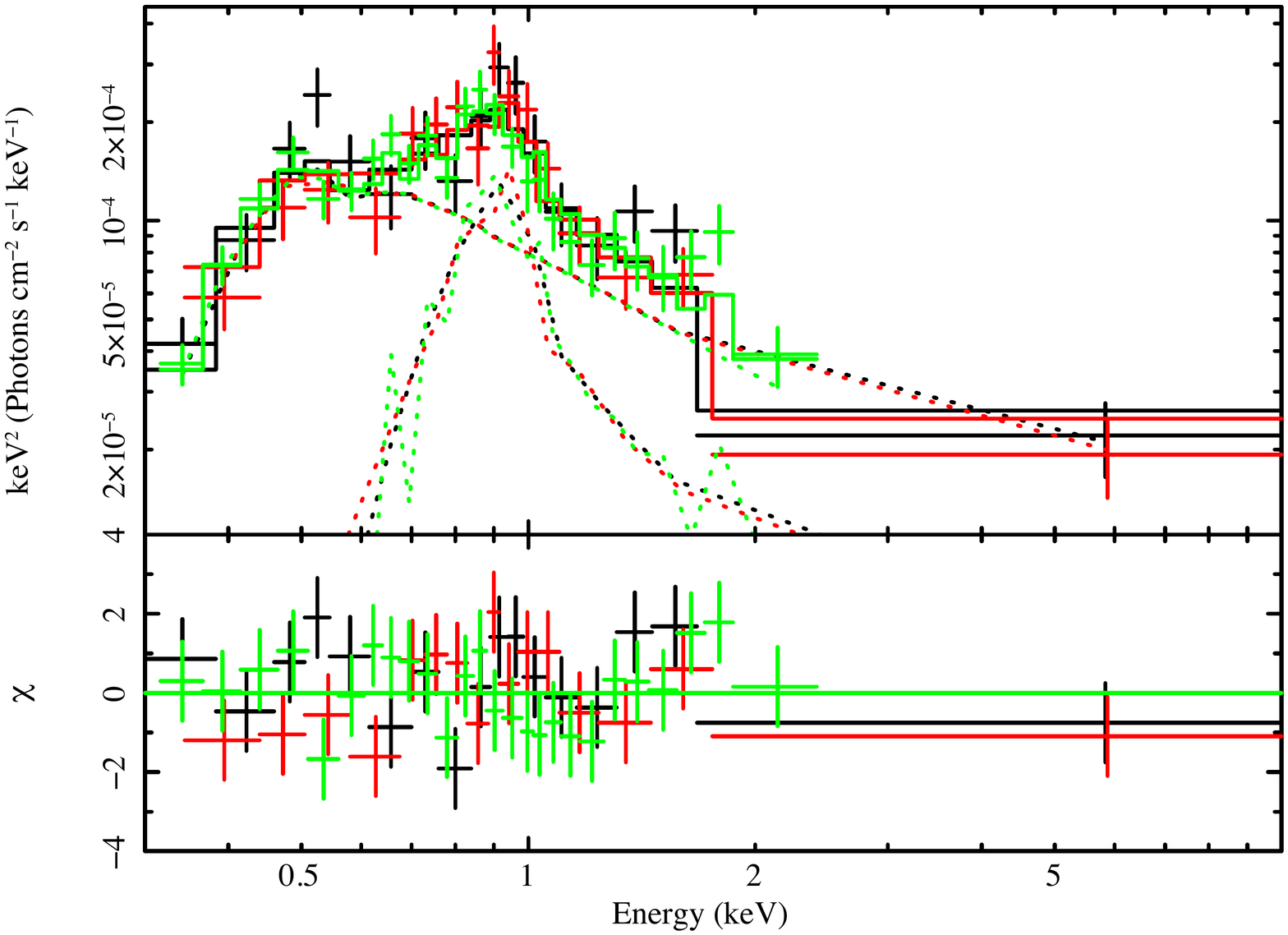}
\vspace{-0.7cm}
\end{center}
\caption{The spectral fitting process, showing the unfolded spectrum and $\Delta\chi^2$ residuals for various models fitted to \xmm~observation X2 of NGC~4395~ULX-1. {\it Top left}, fitted with a single absorbed PL model, with $N_{\rm H}=(29\pm5)\times10^{20}$\,cm$^{-2}$ and $\Gamma=4.3\pm0.3$, $\chi^2=96.3/56$. {\it Top right}, fitted with an absorbed PL model with $N_{\rm H}=12^{+6}_{-5}\times10^{20}$\,cm$^{-2}$ and $\Gamma=3.7\pm0.4$, and a MEKAL thermal plasma component with $kT=0.69^{+0.08}_{-0.07}$\,keV, $\chi^2=61.0/54$. {\it Bottom left}, fitted with an absorbed MCD with $N_{\rm H}=17^{+10}_{-8}\times10^{20}$\,cm$^{-2}$ and $T_{\rm in}=0.1^{+0.2}_{-0.1}$\,keV, a PL with $\Gamma=3.5^{+0.6}_{-0.5}$ and a MEKAL component with $kT=0.69^{+0.09}_{-0.07}$\,keV, $\chi^2=59.6/52$. {\it Bottom right}, fitted with an absorbed MCD with $N_{\rm H}=13^{+20}_{-6}\times10^{20}$\,cm$^{-2}$ and $T_{\rm in}=0.1\pm0.06$\,keV, convolved with a {\tt simpl} convolution with $\Gamma=3.5^{+0.4}_{-0.5}$ and scattered fraction $>0.05$, and a MEKAL component with $kT=0.70\pm0.08$\,keV, $\chi^2=59.5/52$.}
\vspace{-0.4cm}
\label{fig:singleobs}
\end{figure*}

In order to examine potential optical counterparts for our sources, we used pre-processed drizzled data from the Hubble Legacy Archive or the MAST distribution centre for all sources. So that we could match the \chandra~source positions with \hst~images, we refined the \hst~astrometry by aligning the WCS with sources from the USNO 2.0 catalogue using the IRAF tools {\sc ccfind}, {\sc ccmap} and {\sc ccsetwcs}. We found the 1$\sigma$ rms error on the astrometric corrections, converted these errors into 90\% confidence intervals, and combined them in quadrature with the 90\% confidence interval for the overall astrometric accuracy of \chandra~(0.6\,arcsec). This gave us error circles of radius 0.8\,arcsec for NGC~300~X-1, and 0.7\,arcsec for M51~ULS, NGC~4395~ULX-1 and NGC~6946~ULX-1 (rounded to the nearest 0.1\,arcsec). In the cases of M51 and NGC~6946, we found a small number of direct \chandra~and \hst~coincidences (three background AGNs and a foreground star, and three foreground stars respectively). We used these to calculate alternate astrometric corrections and 90\% confidence regions -- 0.7\,arcsec radius for both -- and kept the results of both of these methods for comparison.

We performed aperture photometry with {\sc gaia} on the NGC~6946~ULX-1 counterpart we identify, as it is the only source in our sample without a previous investigation into possible optical counterparts. We used the aperture correction method and STMAG zero-points given in \citet{sirianni05}, and corrected for foreground extinction using the $E(B-V)$ value given in Table~\ref{tab:sources}, with extinction ratios from \citet{sirianni05} assuming an O5 SED.

We show the \xmm~EPIC-pn images and corresponding astrometrically corrected \hst~images for all four sources in Fig.~\ref{fig:images}, with solid circles on the \hst~images representing the error region from correcting the astrometry by aligning with USNO sources, and dashed sources the error region obtained by matching \hst~and \chandra~directly. We retrieve most previously identified optical counterparts to these sources and identify a potential counterpart for NGC~6946~ULX-1 for the first time.

\section{Results}
\label{sec:results}

\subsection{X-ray Spectral Fitting}
\label{sec:spec}

\begin{table*}
\begin{minipage}{172mm}
\caption{The best-fitting parameters of a single-component (MCD or PL) model fit to our sample of sources.} \label{tab:fit1}
\begin{center}
\begin{tabular}{lcccccccc}
  \hline
  Name & ID$^a$ & \multicolumn{3}{c}{{\tt tbabs*tbabs*diskbb}} & \multicolumn{3}{c}{{\tt tbabs*tbabs*powerlaw}}  & \\
   & & $N_{\rm H}^b$ & $T_{\rm in}^c$ & $\chi^2$/d.o.f. & $N_{\rm H}^b$ & $\Gamma^d$ & $\chi^2$/dof & $L_{\rm X}^e$ \\
   & & ($\times10^{20}$\,cm$^{-2}$) & (keV) & & ($\times10^{20}$\,cm$^{-2}$) & & & ($\times10^{38}$\,erg\,s$^{-1}$) \\
  \hline
  NGC~300~X-1 & X1 & 0 & (0.4) & 1057.7/170 & $\mathbf{4\pm1}$ & $\mathbf{2.52\pm0.06}$ & {\bf 227.1/170} & $2.01^{+0.05}_{-0.06}$ \\
   & X2 & 0 & (0.3) & 541.9/101 & 0 & $\mathbf{2.57\pm0.06}$ & {\bf 117.1/101} & $0.86\pm0.02$ \\
   & X3 & 0 & (0.4) & 987.0/166 & $5\pm1$ & $2.60\pm0.06$ & 278.2/166 & $2.15\pm0.05$ \\
   & X4 & 0 & (0.4) & 770.6/149 & $3\pm1$ & $2.55^{+0.08}_{-0.07}$ & 205.0/149 & $1.75\pm0.04$ \\ 
   & X5* & 0 & (0.3) & 241.0/67 & $\mathbf{3\pm2}$ & $\mathbf{2.7\pm0.1}$ & {\bf 83.2/67} & $1.60^{+0.07}_{-0.08}$ \\ 
   & C1 & 0 & $0.46^{+0.06}_{-0.05}$ & 66.1/27 & $\mathbf{8\pm7}$ & $\mathbf{2.6\pm0.3}$ & {\bf 31.9/27} & $1.7^{+0.3}_{-0.2}$ \\ 
   & C2 & 0 & (0.6) & 264.6/86 & $\mathbf{<4.6}$ & $\mathbf{2.4\pm0.1}$ & {\bf 69.4/86} & $1.9^{+0.1}_{-0.2}$ \\
   & C3 & 0 & $0.77\pm0.05$ & 228.2/89 & $\mathbf{8^{+6}_{-5}}$ & $\mathbf{2.5\pm0.1}$ & {\bf 99.0/89} & $2.0\pm0.1$ \\
   & C4 & 0 & $0.78\pm0.06$ & 166.9/63 & $\mathbf{<7.7}$ & $\mathbf{2.3\pm0.1}$ & {\bf 67.2/63} & $2.01^{+0.1}_{-0.3}$ \\\\ 
  NGC~4395~ULX-1 & X1 & $4\pm2$ & $0.29\pm0.02$ & 151.2/83 & $28\pm4$ & $4.2\pm0.2$ & 123.6/83 & $6.1\pm0.3$ \\
   & X2 & $3^{+3}_{-2}$ & $0.28\pm0.02$ & 109.3/56 & $29\pm5$ & $4.3\pm0.3$ & 96.3/56 & $6.6\pm0.3$ \\
   & X3 & (0.5) & (0.3) & 913.4/187 & (0.2) & (4.0) & 649.4/187 & $(7.4)$ \\ 
   & X4** & $6\pm2$ & $0.27\pm0.01$ & 203.7/79 & $33\pm4$ & $4.5\pm0.2$ & 178.7/79 & $8.1\pm0.2$ \\ 
   & C1 & 0 & $\mathbf{0.48\pm0.04}$ & {\bf 29.5/27} & $\mathbf{50\pm30}$ & $\mathbf{3.8\pm0.4}$ & {\bf 19.8/27} & $12.2^{+2}_{-1}$ \\\\ 
  M51~ULS & X1*** & $\mathbf{<8}$ & $\mathbf{0.16^{+0.03}_{-0.02}}$ & {\bf 25.4/16} & $30^{+2}_{-1}$ & $5.8^{+1.2}_{0.8}$ & 33.5/16 & $8.0^{+1.0}_{-0.9}$ \\
   & X2 & $7\pm3$ & $0.15\pm0.01$ & 107.8/57 & $30^{+7}_{-6}$ & $6.1\pm0.5$ & 152.9/57 & $9.0\pm0.4$ \\
   & X3 & $\mathbf{<13}$ & $\mathbf{0.24^{+0.04}_{-0.06}}$ & {\bf 37.8/26} & $\mathbf{20^{+20}_{-10}}$ & $\mathbf{4\pm1}$ & {\bf 34.6/26} & $1.6\pm0.2$ \\ 
   & X4 & $\mathbf{3\pm3}$ & $\mathbf{0.16\pm0.01}$ & {\bf 45.6/40} & $\mathbf{23^{+6}_{-5}}$ & $\mathbf{5.7\pm0.5}$ & {\bf 58.4/40} & $6.6\pm0.3$ \\ 
   & C1 & $10^{+6}_{-5}$ & $0.16^{+0.02}_{-0.01}$ & 58.0/29 & $60^{+20}_{-10}$ & $7\pm1$ & 85.4/29 & $6.5^{+0.5}_{-0.6}$ \\
   & C2 & $10^{+8}_{-6}$ & $0.14\pm0.01$ & 95.1/38 & $70\pm20$ & $9\pm1$ & 110.4/38 & $4.4\pm0.5$ \\
   & C3 & $12^{+9}_{-7}$ & $0.12\pm0.01$ & 78.5/34 & (58) & (8.8) & 105.5/34 & $3.6\pm0.4$ \\ 
   & C4 & $17^{+6}_{-5}$ & $0.14\pm0.01$ & 163.5/56 & (74) & (8.8) & 173.9/56 & $6.2\pm0.3$ \\ 
   & C5 & $20\pm10$ & $0.13\pm0.02$ & 46.2/20 & (85) & (9.4) & 62.6/20 & $5.2\pm0.6$ \\\\
  NGC~6946~ULX-1 & X1 & 0 & $0.28\pm0.03$ & 35.9/15 & $\mathbf{30\pm20}$ & $\mathbf{4.5^{+0.9}_{-0.7}}$ & {\bf 31.4/15} & $3.8^{+0.5}_{-0.4}$ \\
   & X2 & $\mathbf{<4.6}$ & $\mathbf{0.34^{+0.02}_{-0.03}}$ & {\bf 98.1/69} & $\mathbf{38^{+7}_{-6}}$ & $\mathbf{4.3\pm0.3}$ & {\bf 71.3/69} & $6.2\pm0.3$ \\ 
   & X3 & $<0.6$ & $0.47^{+0.02}_{-0.01}$ & 235.7/121 & $\mathbf{29\pm3}$ & $\mathbf{3.4\pm0.1}$ & {\bf 132.8/121} & $13.6\pm0.3$ \\
   & X4 & $(5)$ & $(0.3)$ & 515.0/140 & $4.0^{+0.4}_{-0.3}$ & $4.5\pm0.2$ & 349.8/140 & $5.1\pm0.1$ \\ 
   & C1 & $11\pm5$ & $0.31\pm0.03$ & 131.0/64 & $\mathbf{51\pm7}$ & $\mathbf{4.7\pm0.3}$ & {\bf 88.7/64} & $5.6^{+0.2}_{-0.3}$ \\
   & C2 & $9^{+9}_{-8}$ & $0.30^{+0.05}_{-0.04}$ & 84.3/33 & $\mathbf{50\pm10}$ & $\mathbf{4.7^{+0.6}_{-0.5}}$ & {\bf 63.9/33} & $5.3\pm0.4$ \\ 
   & C3 & $<13$ & $0.32^{+0.02}_{-0.06}$ & 41.3/19 & $\mathbf{40\pm20}$ & $\mathbf{4.6^{+0.9}_{-0.8}}$ & {\bf 35.5/19} & $3.5\pm0.4$ \\
   & C4 & (37) & (0.2) & 50.1/16 & $80^{+30}_{-20}$ & $6\pm1$ & 40.0/16 & $2.8^{+0.4}_{-0.3}$ \\
   & C5 & $\mathbf{20^{+20}_{-10}}$ & $\mathbf{0.27^{+0.06}_{-0.05}}$ & {\bf 29.3/15} & $\mathbf{70^{+30}_{-20}}$ & $\mathbf{5.5^{+1.1}_{-0.9}}$ & {\bf 19.3/15} & $2.8^{+0.4}_{-0.2}$ \\ 
  \hline
\end{tabular}
\end{center}
\vspace{-2mm}
Notes: Statistically acceptable fits (confidence within 3$\sigma$ based on $\chi^2$ statistic) are displayed in bold. Values given in brackets are not constrained due to a poor fit. *M2 data was not used due to normalisation issues. **No PN data available due to the source being very off-axis. ***PN data was not used due to normalisation issues as the source was close to a chip gap. \\
$^a$The short observation ID given in Table~\ref{tab:observations}. \\
$^b$The intrinsic column density of the source. The Galactic column density is accounted for in the first, frozen {\tt tbabs} component; see Table~\ref{tab:sources}. Models for which no further absorption component was required are indicated with $N_H=0$. \\
$^c$The accretion disc temperature at the inner disc radius. \\
$^d$The photon index of the power-law model. \\
$^e$The source luminosity calculated using the best-fitting single-component model (a MCD for M51~ULS and a PL for the other sources) from the observed flux between 0.3 and 10\,keV.
\vspace{-0.3cm}
\end{minipage}
\end{table*}

Spectral fitting was performed using {\sc v12} of XSPEC \citep{xspec} in the energy range of 0.3--10\,keV, with errors calculated at the 90\% confidence interval. The best-fitting model parameters were found using $\chi^2$ minimisation, and $\chi^2$ statistics were used to determine the goodness-of-fit. We used the abundance tables of \citet{wilms00} throughout. We began with simple single-component models and refined the models if necessary as described in Sections~\ref{sec:resid} and~\ref{sec:phys}. All models also contain an absorption component frozen to the Galactic column (see Table~\ref{tab:sources}). An example of the spectral fitting process is shown in Fig.~\ref{fig:singleobs}. 

\subsubsection{Single-component Models}
\label{sec:singcomp}

We began by fitting the source spectra with absorbed single-component models, either a multicolour disc (MCD; {\tt tbabs*diskbb}) or a power-law (PL; {\tt tbabs*powerlaw}).

We present the results of a first pass of single-component fitting in Table~\ref{tab:fit1}. The first thing we find is that all four sources are very soft, with MCD models taking on low temperatures and PL models having high photon indices. However, only M51~ULS is better fitted by a MCD model than by a PL, with three of its \xmm~observations acceptably fit by a MCD without any additional components in the model, although additional spectral features are seen in the residuals of most of the observations.

\begin{table*}
\begin{minipage}{165mm}
\caption{The best-fitting parameters of multi-component (a MCD or PL component plus one or more residual components) model fits, for observations for which a refined model offered a significant improvement to the goodness-of-fit (i.e. $\Delta\chi^2 > 10/$d.o.f.). MCD and PL parameters as in Table~\ref{tab:fit1}.} \label{tab:fit2}
\begin{center}
\begin{tabular}{lc@{~}c@{~}c@{~}c@{~~}c@{~}c@{~~}c@{~}c@{~~}c@{~~}c@{~~}c}
  \hline
  Name & ID & \multicolumn{2}{c}{{\tt tbabs*tbabs*(diskbb}} & \multicolumn{2}{c}{{\tt *edge}} & {\tt +mekal} & {\tt +mekal} & \multicolumn{2}{c}{{\tt +gauss}} & ) & \\
   & & $N_{\rm H}$ & $T_{\rm in}$ & $E_c^a$ & $\tau^b$ & $kT_1^c$ & $kT_2^c$ & $E_L^d$ & $\sigma_L^e$ & $F^f$ & $\Delta\chi^2/\Delta$dof \\
   & & ($\times10^{20}$\,cm$^{-2}$) & (keV) & (keV) &  & (keV) & (keV) & (keV) & (keV) & & \\
  \hline
  M51~ULS & X2 & $6^{+6}_{-4}$ & $0.12\pm0.02$ & ... & ...& $0.50^{+0.08}_{-0.12}$ & ... & ... & ...& 0.34 & 29.7/2 \\ 
   & & $<3$ & $0.21^{+0.01}_{-0.02}$ & $0.97\pm0.03$ & $1.7\pm0.6$ & ... & ... & ... & ... & ... & 33.7/2 \\
   & C1 & $<4$ & $0.22\pm0.03$ & $1.10^{+0.05}_{-0.04}$ & $1.7^{+0.7}_{-0.6}$ & ... & ... & ... & ... & ... & 24.4/2 \\
   & C2 & $<12$ & $0.11\pm0.02$ & ... & ... & $0.55^{+0.04}_{-0.08}$ & ... & ... & ... & 0.30 & 33.5/2 \\
   & & $<5$ & $0.20^{+0.01}_{-0.02}$ & $0.99\pm0.02$ & $1.8^{+0.6}_{-0.5}$ & ... & ... & ... & ... & ... & 41.1/2 \\
   & C3 & 0 & $0.18^{+0.02}_{-0.01}$ & $0.95\pm0.03$ & $2.1^{+0.8}_{-0.5}$ & ... & ... & ... & ... & ... & 38.2/2 \\
   & C4 & $16^{+10}_{-8}$ & $0.10\pm0.01$ & ... & ... & $0.59\pm0.05$ & ... & ... & ... & 0.15 & 76.8/2 \\ 
   & & $3^{+4}_{-3}$ & $0.21\pm0.02$ & $1.04\pm0.02$ & $1.8\pm0.4$ & ... & ... & ... & ... & ... & 83.0/2 \\
   & C5 & 0 & $0.23\pm0.04$ & $1.06\pm0.03$ & $2.0^{+0.6}_{-0.5}$ & ... & ... & ... & ... & ... & 23.2/2 \\ 
  \hline
  Name & ID & \multicolumn{2}{c}{{\tt tbabs*tbabs*(powerlaw}} & \multicolumn{2}{c}{{\tt *edge}} & {\tt +mekal} & {\tt +mekal} & \multicolumn{2}{c}{{\tt +gauss}} & ) & \\
   & & $N_{\rm H}$ & $\Gamma$ & $E_c^a$ & $\tau^b$ & $kT_1^c$ & $kT_2^c$ & $E_L^d$ & $\sigma_L^e$ & $F^f$ & $\Delta\chi^2/\Delta$dof \\
   & & ($\times10^{20}$\,cm$^{-2}$) &  & (keV) &  & (keV) & (keV) & (keV) & (keV) & & \\
  \hline
  NGC~300~X-1 & X1 & $2\pm1$ & $2.42\pm0.06$ & ... & ... & ... & ... & $0.93^{+0.02}_{-0.03}$ & $0.07\pm0.03$ & 0.03 & 56.6/3 \\
   & X2 & 0 & $3.1\pm0.2$ & ... & ... & $(12.2)$ & ... & ... & ... & 0.50 & 41.4/2 \\
   & X3 & $2\pm1$ & $2.46\pm0.07$ & ... & ... & ... & ... & $0.93\pm0.02$ & $0.08\pm0.02$ & 0.05 & 76.9/3 \\
   & X4 & $<1$ & $2.41^{+0.08}_{-0.06}$ & ... & ... & ... & ... & $0.90^{+0.02}_{-0.03}$ & $0.07\pm0.02$ & 0.03 & 58.6/3 \\ 
   & X5 & $3\pm2$ & $2.5\pm0.1$ & $1.39\pm0.05$ & $0.6\pm0.2$ & ... & ... & ... & ... & ... & 26.3/2 \\\\
  NGC~4395~ULX-1 & X1 & $20\pm5$ & $3.9^{+0.4}_{-0.3}$ & ... & ... & $0.8\pm0.1$ & ... & ... & ... & 0.14 & 20.9/2 \\
   & X2 & $12^{+6}_{-5}$ & $3.7\pm0.4$ & ... & ... & $0.69^{+0.08}_{-0.07}$ & ... & ... & ... & 0.42 & 35.3/2 \\
   & X3* & $14\pm1$ & $3.63\pm0.08$ & ... & ... & $0.75\pm0.03$ & ... & ... & ... & 0.34 & 379.2/2 \\ 
   & & $13^{+2}_{-1}$ & $3.8^{+0.2}_{-0.1}$ & ... & ... & $0.68\pm0.03$ & $1.5^{+0.3}_{-0.2}$ & ... & ... & 0.38 & 397.9/4 \\ 
   & X4 & $21^{+5}_{-4}$ & $4.1\pm0.3$ & ... & ... & $0.77^{+0.10}_{-0.06}$ & ... & ... & ... & 0.29 & 68.5/2 \\\\ 
  NGC~6946~ULX-1 & X4 & $20\pm4$ & $3.6\pm0.2$ & ... & ... & ... & ... & $0.92^{+0.02}_{-0.03}$ & $0.13\pm0.02$ & 0.34 & 182.8/3 \\
   & C1 & $54\pm7$ & $4.6\pm0.3$ & $1.20^{+0.09}_{-0.05}$ & $0.5\pm0.2$ & ... & ... & ... & ... & ... & 21.3/2 \\ 
   & C2 & $20\pm10$ & $3.3^{+0.7}_{-0.6}$ & ... & ... & $0.7\pm0.1$ & ... & ... & ... & 0.36 & 22.7/2 \\ 
   & & $50\pm10$ & $4.5^{+0.5}_{-0.4}$ & $1.17\pm0.04$ & $0.8\pm0.3$ & ... & ... & ... & ... & ... & 20.0/2 \\ 
   & C4 & $70^{+60}_{-50}$ & $(7.7)$ & ... & ... & $0.8\pm0.2$ & ... & ... & ... & 0.72 & 22.2/2 \\ 
   & & $80^{+30}_{-20}$ & $5.3^{+1.0}_{-0.8}$ & $1.18^{+0.07}_{-0.05}$ & $1.3^{+0.6}_{-0.5}$ & ... & ... & ... & ... & ... & 20.0/2 \\ 
  \hline
\end{tabular}
\end{center}
\vspace{-2mm}
Notes: \xmm~observations with particular cameras excluded are the same as in Table~\ref{tab:fit1}. *Both refined fits of X3 are major improvements on a single-component fit, but are not in themselves statistically acceptable.\\
$^a$Threshold energy of absorption edge. \\
$^b$Absorption edge depth. \\
$^c$The plasma temperature of a first and, optionally, second MEKAL component. \\
$^d$Energy of Gaussian emission line. \\
$^e$Line width of Gaussian emission line.\\
$^f$Flux fraction of the residual component(s).
\vspace{-0.3cm}
\end{minipage}
\end{table*}

The majority of observations of both NGC~300~X-1 and NGC~6946~ULX-1 show very significant hard residuals when fitted with a MCD, but are well-fitted by a PL model, with photon index $\Gamma\sim2.5$ for NGC~300~X-1 and a high photon index $\Gamma\sim3.5-5.5$ for NGC~6946~ULX-1. Examination of the residuals of the NGC~300~X-1 spectra, especially those for which a single-component model was not an adequate fit, reveal that there is a peaked excess around 0.9\,keV in the \xmm~data. This is consistent with the findings of \citet{carpano07} who find that the inclusion of a Gaussian line at that energy significantly improves the spectral fit. Observation X4 of NGC~6946~ULX-1 also shows distinctive soft residuals around 1\,keV as seen in other ULX spectra (e.g. \citealt{middleton15b}). We therefore refined our models for NGC~300~X-1 and NGC~6946~ULX-1 to include an additional broad Gaussian line to fit the excess at $\sim0.9$\,keV if it offered a significant improvement to the fit. For other observations of NGC~6946~ULX-1, while well-fitted by a single-component PL model, many of them still exhibit residuals potentially indicative of hot gas emission or absorption edges.

The observations for NGC~4395~ULX-1 were fitted better by a PL model than by a MCD, but unlike NGC~300~X-1 and NGC~6946~ULX-1, only the observation with the lowest amount of data was acceptably fit with a single-component model. The other observations exhibited strong residuals suggesting the presence of a thermal plasma component.

\subsubsection{Fitting the residuals}
\label{sec:resid}

Since many of the observations showed significant residuals when fitted with single-component models, we next attempted to fit these residuals by adding a {\tt mekal} or {\tt edge} component to the model to test for contributions from thermal plasma or ionised absorption features. In the case of NGC~300~X-1, we used a Gaussian line at $\sim0.9$\,keV as in \citet{carpano07}. The observations for which the fit was significantly improved ($\Delta\chi^2>10$ per d.o.f. over the original single-component fit) by the addition of such a component are given in Table~\ref{tab:fit2}.

\begin{figure*}
\begin{center}
\vspace{-0.5cm}
\includegraphics[width=9cm]{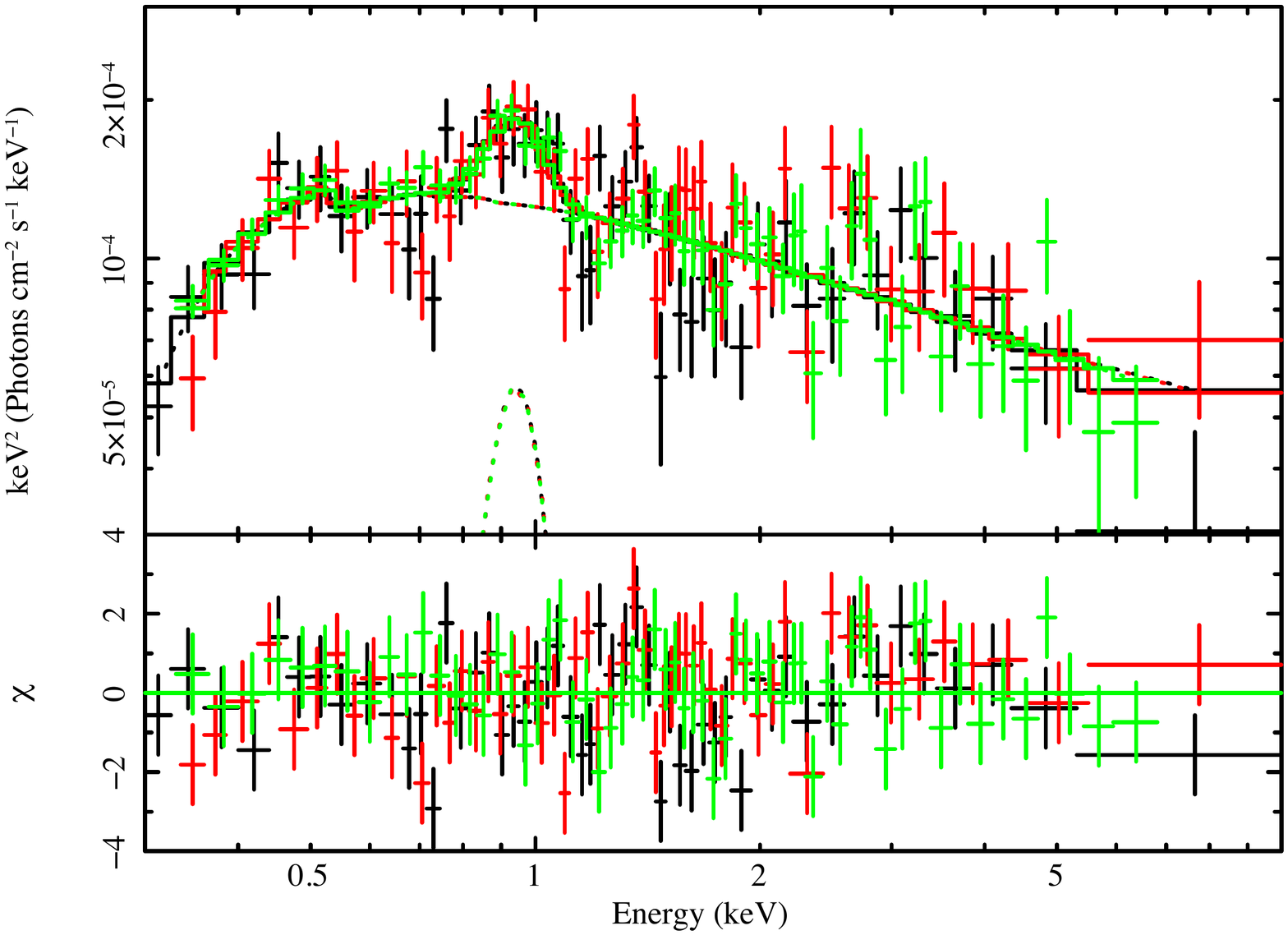}\hspace{-0.5cm}
\includegraphics[width=9cm]{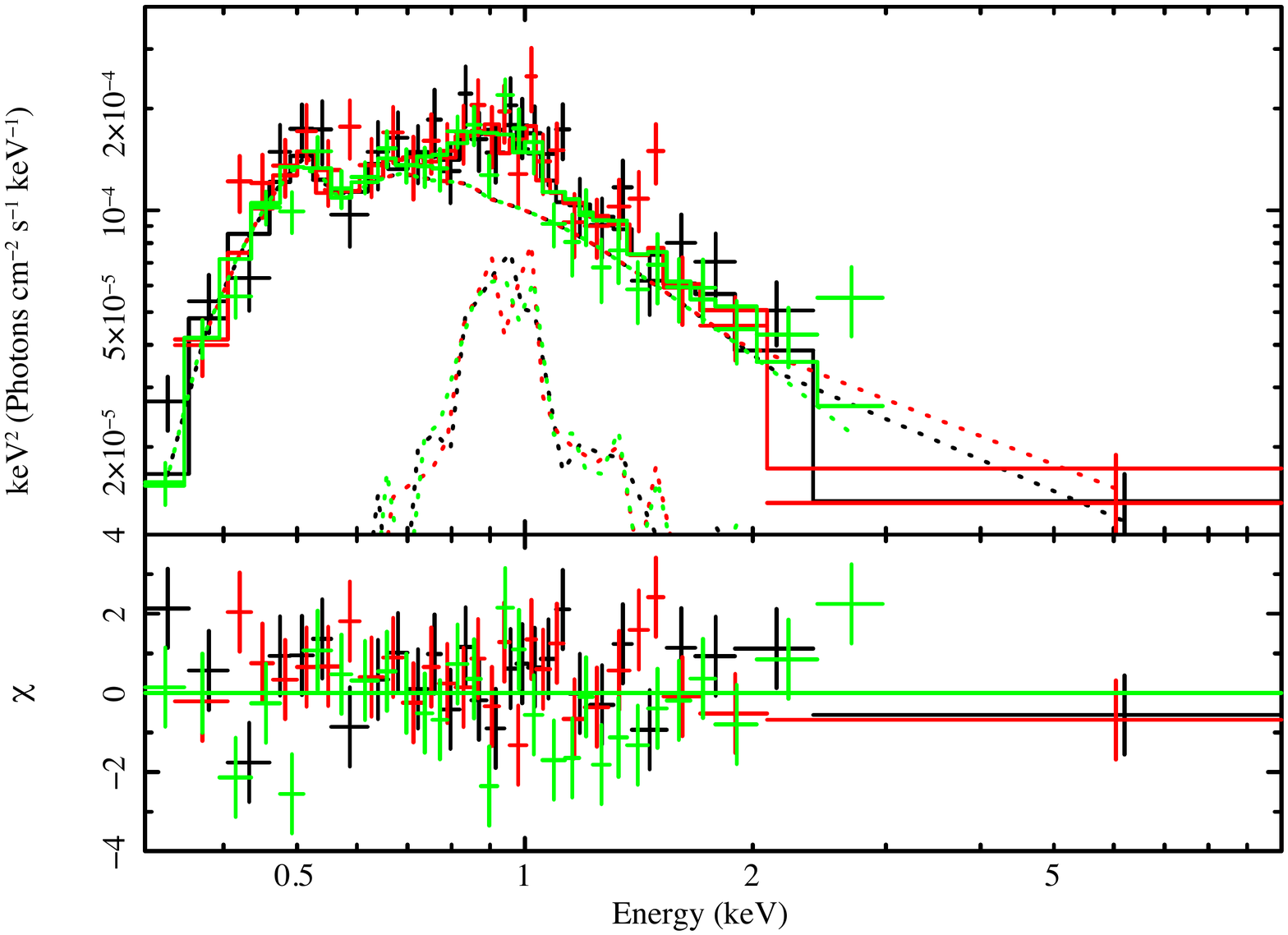}\vspace{-1cm}
\includegraphics[width=9cm]{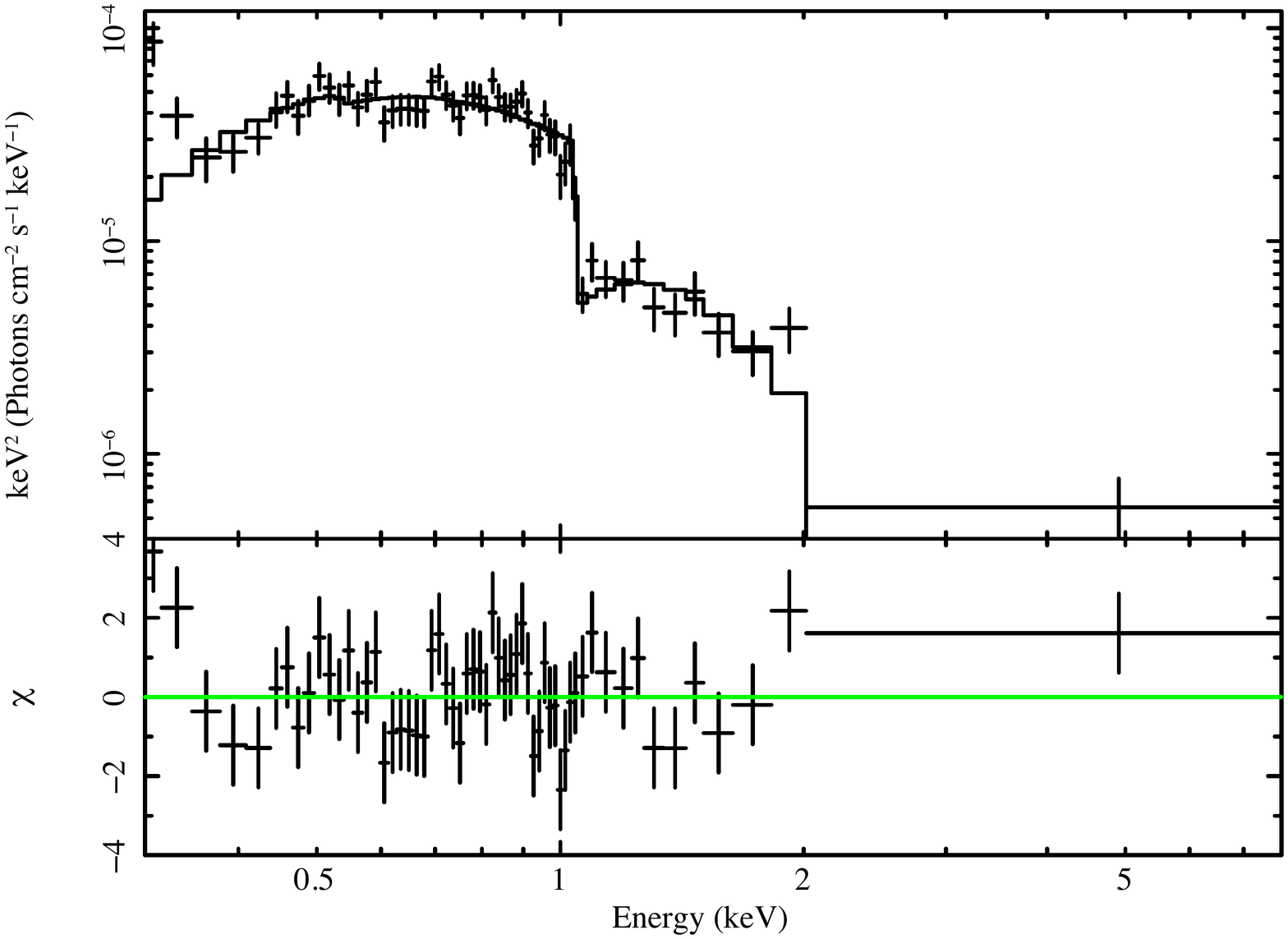}\hspace{-0.5cm}
\includegraphics[width=9cm]{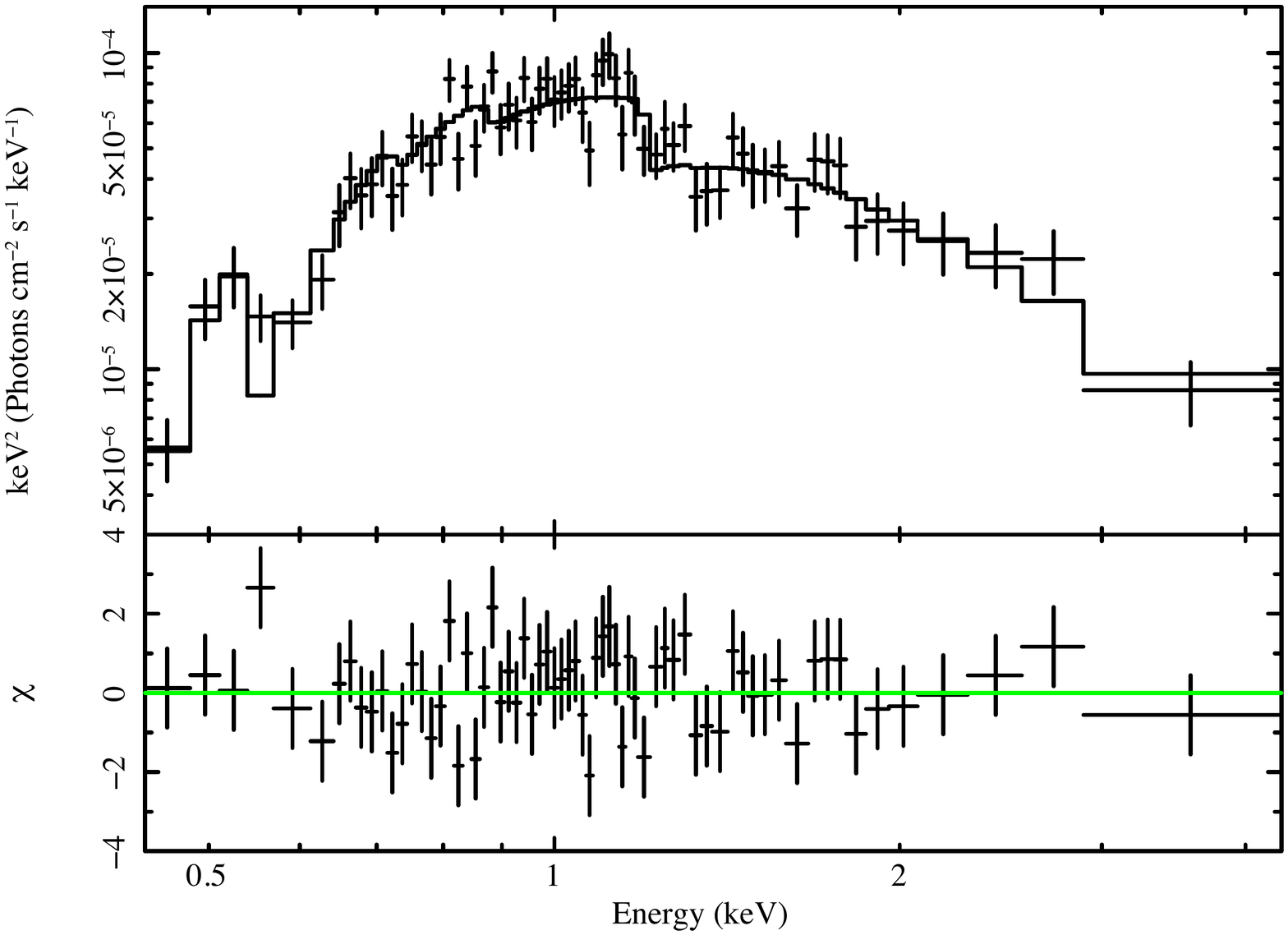}
\vspace{-0.5cm}
\end{center}
\caption{The unfolded spectrum and $\Delta\chi^2$ residuals for the best-fitting absorbed single component plus residuals model, for a typical observation of each source. {\it Top left}, NGC~300~X-1 observation X3 fitted with an absorbed PL model with $N_{\rm H}=2\pm1\times10^{20}$\,cm$^{-2}$ and $\Gamma=2.46\pm0.07$, and an additional Gaussian line with $E_L=0.93\pm0.02$\,keV and $\sigma_L=0.08\pm0.02$\,keV, $\chi^2=201.3/163$. {\it Top right}, NGC~4395~ULX-1 observation X1, fitted with an absorbed PL model with $N_{\rm H}=20\pm5\times10^{20}$\,cm$^{-2}$ and $\Gamma=3.9^{+0.4}_{-0.3}$, and an additional MEKAL component with $kT=0.8\pm0.1$\,keV, $\chi^2=102.7/81$. {\it Bottom left}, M51~ULS observation C4, fitted with an absorbed MCD model with $N_{\rm H}=3^{+4}_{-3}\times10^{20}$\,cm$^{-2}$ and $T_{\rm in}=0.21\pm0.02$\,keV, with an absorption edge at $E_c=1.04\pm0.02$\,keV with depth $\tau=1.8\pm0.4$, $\chi^2=80.5/54$. {\it Bottom right}, NGC~6946~ULX-1 observation C1, fitted with an absorbed PL model with $N_{\rm H}=54\pm7\times10^{20}$\,cm$^{-2}$ and $\Gamma=4.6\pm0.3$, with an absorption edge at $E_c=1.20^{+0.09}_{-0.05}$\,keV with depth $\tau=0.5\pm0.2$, $\chi^2=67.4/62$.}
\label{fig:typobs}
\end{figure*}

In the case of M51~ULS, the MCD fit can be improved either by the addition of a $\sim0.5$\,keV thermal plasma or a $\sim1$\,keV absorption edge, the latter offering slightly better improvement to the reduced $\chi^2$ statistic. Including an absorption edge results in a slightly higher temperature for the underlying disc, at around 0.2\,keV compared to around 0.1\,keV in the case of adding a {\tt mekal} component.

Three of the \xmm~observations of NGC~300~X-1 are significantly improved by the addition of a broad Gaussian line at $\sim0.9$\,keV. One exception was observation X2, which showed curvature in its residuals which we were able to fit with an ill-constrained high-temperature {\tt mekal} component, but is likely more indicative of there being two underlying components to its emission similar to the canonical BH accretion states (see Section~\ref{sec:phys}). Observation X5 was also not improved by the addition of a Gaussian line, instead exhibiting an absorption feature at $E_c=1.39\pm0.05$\,keV which is unique to that observation. The \chandra~observations, on the other hand, showed no evidence of additional spectral features on top of the underlying PL.

All \xmm~observations of NGC~4395~ULX-1 are improved by the addition of a 0.7--0.8\,keV {\tt mekal} component. In particular, the fit to observation X3 is dramatically improved by the addition of the thermal plasma component, but is still statistically unacceptable, with null hypothesis probability $p=4.4\times10^{-5}$. The fit can be improved further, with diminishing returns, by adding more {\tt mekal} components, but the residuals are too complex to be well-described by this method.

While most observations of NGC~6946~ULX-1 do not undergo significant improvements to their fit with the inclusion of additional components, three of the \chandra~observations appear to exhibit an absorption feature at $\sim1.2$\,keV. In two of these cases, the spectrum can also be fitted with a warm {\tt mekal} component, although in the case of C4, it has the effect of dramatically steepening the PL to the point that it becomes very unconstrained, and the thermal plasma component dominates the emission. Observation X4 exhibits strong soft residuals that are not well-described either by an absorption edge or {\tt mekal} components, and instead resemble residual peaks at $\sim$1\,keV that are found in ULX spectra, therefore we fit these with a Gaussian component (we also note that two Gaussian absorption components can be used to produce a similar improvement in fit as in \citealt{middleton15b}, but for simplicity we only use a single Gaussian emission component in our analysis). 

We show a typical example of a spectrum of each source fitted with a single underlying component plus residuals in Fig.~\ref{fig:typobs}.

\subsubsection{Towards physically motivated models}
\label{sec:phys}

\begin{table*}
\begin{minipage}{172mm}
\caption{The best-fitting parameters of physically-motivated fits to the spectra with an underlying PL shape. Parameters as in Table~\ref{tab:fit1}.} \label{tab:fit3}
\begin{center}
\begin{tabular}{lccccccccc}
  \hline
  Name & ID$^a$ & \multicolumn{2}{c}{{\tt tbabs*tbabs*diskbb}} & \multicolumn{2}{c}{{\tt *simpl}} & \multicolumn{2}{c}{{\tt +comptt}} & {\tt )} & \\
   & & $N_{\rm H}$ & $T_{\rm in}$ & $\Gamma$ & $F_{\rm scat}^a$ & $kT^b$ & $\tau^c$ & $\chi^2$/dof & R$^d$ \\
   & & ($\times10^{20}$\,cm$^{-2}$) & (keV) & & & (keV) & & \\
  \hline
  NGC~300~X-1 & X1 & $7^{+5}_{-4}$ & $0.09^{+0.02}_{-0.01}$ & $2.44\pm0.08$ & $0.3^{+0.4}_{-0.1}$ & ... & ... & 170.9/165 & G \\ 
   & & $6\pm5$ & $0.08^{+0.01}_{-0.02}$ & ... & ... & $<17.5$ & $5^{+1}_{-3}$ & 165.2/164 & G \\
   & X2 & 0 & $0.16\pm0.03$ & $2.0\pm0.2$ & $0.29\pm0.09$ & ... & ... & 70.7/98 & \\ 
   & X3 & $2^{+4}_{-3}$ & $0.12^{+0.05}_{-0.03}$ & $2.42^{+0.07}_{-0.09}$ & $0.5^{+0.2}_{-0.3}$ & ... & ... & 208.0/161 & G \\ 
   & & $<5$ & $0.10\pm0.03$ & ... & ... & $<33.9$ & $5.4^{+0.4}_{-4.9}$ & 197.1/160 & G \\
   & X4 & $7\pm4$ & $0.10^{+0.03}_{-0.02}$ & $2.3\pm0.1$ & $0.2^{+0.2}_{-0.1}$ & ... & ... & 128.6/144 & G \\ 
   & & $8^{+6}_{-5}$ & $0.10\pm0.02$ & ... & ... & $<9.9$ & $5.6^{+0.4}_{-4.0}$ & 126.1/143 & G \\
   & X5 & 0 & $0.21^{+0.02}_{-0.03}$ & $2.3^{+0.1}_{-0.3}$ & $0.34^{+0.08}_{-0.10}$ & ... & ... & 68.2/63 & E \\
   & C1 & $<12$ & $0.20^{+0.10}_{-0.07}$ & $2.0^{+0.6}_{-0.9}$ & $0.3^{+0.4}_{-0.2}$ & ... & ... & 27.9/25 & \\ 
   & & $<21$ & $0.19^{+0.10}_{-0.07}$ & ... & ... & $<109$ & $<21.4$ & 27.7/24 & \\
   & C2 & $<33$ & $0.17^{+0.07}_{-0.04}$ & $2.3^{+0.1}_{-0.2}$ & $0.4^{+0.3}_{-0.2}$ & ... & ... & 62.8/84 & \\ 
   & & $<15$ & $0.17^{+0.06}_{-0.03}$ & ... & ... & $<28.9$ & $<7.0$ & 60.9/83 & \\
   & C4 & $<75$ & $0.13^{+0.08}_{-0.03}$ & $2.4^{+0.2}_{-0.3}$ & $0.1^{+0.6}_{-0.1}$ & ... & ... & 63.7/61 & \\ 
   & & $<51$ & $0.14^{+0.06}_{-0.03}$ & ... & ... & $<100$ & $5.9^{+0.7}_{-5.0}$ & 60.9/60 & \\\\ 
  NGC~4395~ULX-1 & X1 & $14\pm10$ & $0.12\pm0.05$ & ... & ... & $<130$ & $<4.4$ & 98.1/78 & M \\ 
   & X2 & $14^{+13}_{-9}$ & $0.10^{+0.06}_{-0.08}$ & $3.5^{+0.4}_{-0.5}$ & $>0.05$ & ... & ... & 59.5/52 & M \\
   & X3 & $11\pm1$ & $0.041^{+0.01}_{-0.004}$ & $3.55^{+0.05}_{-0.08}$ & $>0.08$ & ... & ... & 269.5/183 & M \\
   & & $9^{+2}_{-1}$ & $0.090\pm0.001$ & $3.7^{+0.2}_{-0.1}$ & $>0.45$ & ... & ... & 246.1/181 & M+M \\
   & & $8\pm3$ & $0.073^{+0.008}_{-0.2}$ & ... & ... & $<3.5$ & $3.5^{+0.1}_{-1.2}$ & 264.3/182 & M \\
   & & $8\pm3$ & $0.07^{+0.02}_{-0.01}$ & ... & ... & $<3.5$ & $3.3^{+0.2}_{-1.9}$ & 243.4/180 & M+M \\
   & X4 & $13^{+5}_{-4}$ & $0.16\pm0.03$ & $3.6\pm0.8$ & $0.2^{+0.3}_{-0.1}$ & ... & ... & 104.3/75 & M \\
   & & $12^{+5}_{-3}$ & $0.16^{+0.03}_{-0.06}$ & ... & ... & $<135$ & $4^{+1}_{-3}$ & 103.6/74 & M \\ 
   & C1 & $40^{+100}_{-10}$ & $<0.41$ & $3.7^{+0.4}_{-2.7}$ & $>0.22$ & ... & ... & 19.1/25 & \\\\
  NGC~6946~ULX-1 & X1 & $<55$ & $0.17\pm0.09$ & $3^{+2}_{-1}$ & $>0.08$ & ... & ... & 28.6/13 & \\ 
   & X2 & $22^{+20}_{-7}$ & $<0.22$ & $3.9^{+0.4}_{-0.6}$ & $>0.18$ & ... & ... & 68.9/67 & \\ 
   & & $21^{+20}_{-9}$ & $0.15^{+0.06}_{-0.04}$ & ... & ... & $<103$ & $<4.1$ & 68.5/66 & \\
   & X3 & $11^{+4}_{-9}$ & $0.23^{+0.06}_{-0.05}$ & $3.2^{+0.1}_{-0.4}$ & $>0.33$ & ... & ... & 130.5/119 & \\ 
   & & $33^{+7}_{-5}$ & $0.03^{+0.02}_{-0.01}$ & ... & ... & $90^{+10}_{-90}$ & $<0.89$ & 123.6/118 & \\
   & X4 & $60^{+20}_{-30}$ & $0.07\pm0.02$ & $3.8^{+0.2}_{-0.3}$ & $<0.12$ & ... & ... & 159.5/135 & G \\
   & & $50\pm30$ & $0.07\pm0.01$ & ... & ... & $<7.5$ & $3.3^{+0.3}_{-3.0}$ & 158.7/134 & G \\
   & C1 & $11^{+10}_{-4}$ & $0.32^{+0.04}_{-0.02}$ & $>4.1$ & $>0.5$ & ... & ... & 63.8/60 & E \\
   & C2 & $<54$ & $0.2^{+0.1}_{-0.2}$ & $1.8^{+1.6}_{-0.8}$ & $>0.05$ & ... & ... & 39.9/29 & M \\ 
   & C3 & $60^{+40}_{-30}$ & $0.15^{+0.08}_{-0.05}$ & $3\pm2$ & $0.04^{+0.20}_{-0.02}$ & ... & ... & 28.9/17 & \\
  \hline
\end{tabular}
\end{center}
\vspace{-2mm}
Notes: \xmm~observations with particular cameras excluded are the same as in Table~\ref{tab:fit1}. Residual components were used in the fits as appropriate, although their parameters are largely the same as in Table~\ref{tab:fit2}, so we do not display them here. We also only show the fit results for observations where we could obtain minimal constraints for all parameters; observations for which at least one parameter was completely unconstrained are not included. \\
$^a$The fraction of input photons that are scattered. \\
$^b$The plasma temperature. \\
$^c$The plasma optical depth.\\
$^d$The residual component used used in this fit: M = {\tt mekal}, E = {\tt edge}, G = {\tt gauss}.
\end{minipage}
\end{table*}

While a PL model, combined with a characterisation of any complex residuals, is a good empirical description of the spectrum for the sources NGC~300~X-1, NGC~4395~ULX-1 and NGC~6946~ULX-1, it is not in itself suitable for a physical explanation of the system. We therefore attempted to fit these sources with a combined MCD and PL model, for comparison with the canonical BH accretion states defined by an accretion disc and an additional hard Comptonised component.

In the case of \xmm~observations of NGC~300~X-1, the spectra can be fitted using a MCD with $T_{\rm in}\sim0.1$\,keV, a PL with $\Gamma\sim2.4$ and a Gaussian line, for those observations that exhibited the line in their residuals. The other two \xmm~observations are once again exceptions, with X2 exhibiting a harder PL with $\Gamma=2.0\pm0.1$, and X5 exhibiting a hotter accretion disc with $T_{\rm in}=0.24^{+0.03}_{-0.04}$\,keV. Fits to the \chandra~spectra tend to prefer slightly warmer discs of $T_{\rm in}\sim0.2$\,keV, however these are far less well-constrained than in the \xmm~observations. The \chandra~PL photon indices are all consistent with $\Gamma=2.4$ but similarly not well constrained.

NGC~4395~ULX-1 and NGC~6946~ULX-1 can be treated in a similar way, although it is unusual to see PL slopes this steep. To begin with, we tried to fit the spectra with a MCD, using additional {\tt mekal} components to account for the hard excess. We found that subsequent {\tt mekal} components preferentially fitted low-energy residuals and did not account for most of the high-energy tail. Since we were unable to characterise the hard excess with hot diffuse emission, we continued with a traditional combined MCD and PL approach. However, since the PL slopes are still steep ($\Gamma>3$ in most cases) even when a MCD component is included, the PL component dominates the spectral continuum at low energies. This has the effect of potentially artificially reducing the MCD component's temperature and normalisation, and therefore distorting our picture of the physical processes (which could also be happening to a lesser extent to the NGC~300~X-1 spectra, since the PL slopes observed for that source are still fairly steep).

Therefore we instead made use of the {\tt simpl} convolution model component \citep{steiner09}, which self-consistently generates a PL tail from the Compton upscattering of seed photons from an input model, for which we used {\tt diskbb} as before. In the cases where there was sufficient data to put constraints upon the parameters, we also fitted the spectra with a MCD plus {\tt comptt} model, with the input temperature $kT$ tied to the MCD temperature.

\begin{figure*}
\begin{center}
\vspace{-0.5cm}
\includegraphics[width=18cm]{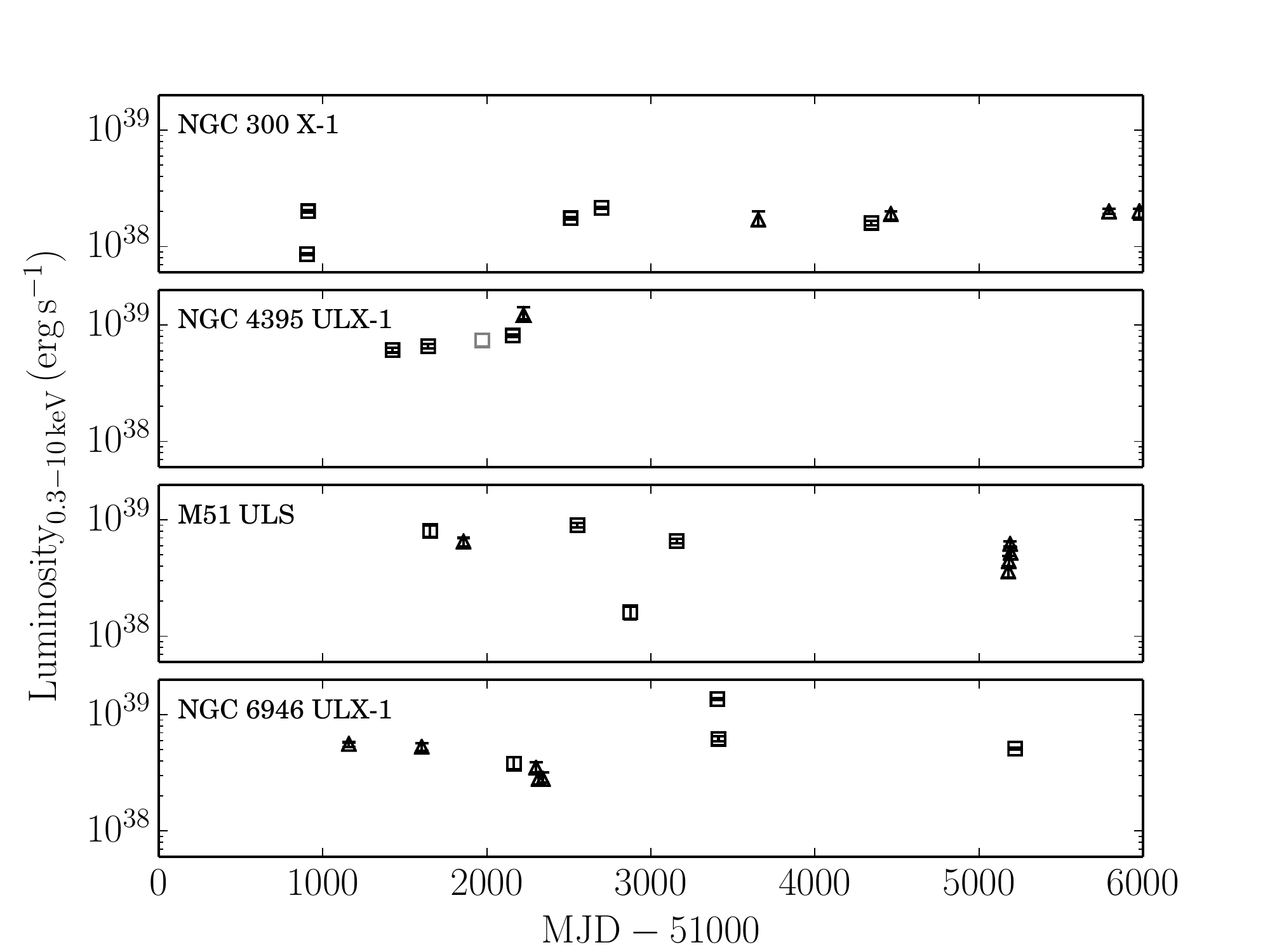}
\end{center}
\vspace{-0.2cm}
\caption{The long term light curve for the sample of Eddington threshold sources using the observed luminosity from \xmm~and \chandra~data, as tabulated in Table~\ref{tab:fit1}. \xmm~observations are marked with squares, \chandra~observations with triangles. The grey square for NGC~4395~ULX-1 without error bars is the observation for which the flux could not be well-constrained.}
\vspace{-0.3cm}
\label{fig:lc}
\end{figure*}

While there is not sufficient data quality to put good constraints on the {\tt simpl} or {\tt comptt} parameters for the majority of observations, we can still glean some overall trends. For all three objects, a cool accretion disc temperature is favoured, with disc temperatures for the most part $<0.2$\,keV. The slopes of the PL tails obtained once the contribution from an accretion disc is properly considered are slightly harder than a PL model on its own, but still unusually soft in the cases of NGC~4395~ULX-1 and NGC~6946~ULX-1. For those observations for which we were able to make a {\tt comptt} fit, the parameters for a Comptonising corona are for the most part very unconstrained, but the majority are consistent with optically thick material and a cool plasma temperature.

\subsection{Timing Analysis}
\label{sec:time}

In order to characterise the long-term variability of each source, we created a long-term light curve (Fig.~\ref{fig:lc}) by using the best-fitting single component model to calculate the 0.3--10\,keV flux of the object in each observation, then using the galaxy distance given in Table~\ref{tab:sources} to calculate the source luminosity (Table~\ref{tab:observations}), assuming $L_{\rm X}=4\pi d^2 f_{\rm X}$ in the absence of direct information on the viewing angle and geometry of the system, where $d$ is the distance to the host galaxy in Mpc, and $f_{\rm X}$ is the X-ray flux between 0.3 and 10\,keV. We do not correct for absorption since extending steep power-laws to low energy might misrepresent (i.e. overestimate) the true luminosity hidden by the absorption -- doing so gives an estimated intrinsic luminosity up to an order of magnitude higher than the observed luminosity, which is most pronounced in the case of NGC~6946~ULX-1.

All four sources do not exhibit large amplitude long-term variability between most of their \xmm~and \chandra~observations. NGC~300~X-1 and M51~ULS both only show one departure from their dominant luminosity which is significantly less bright than the other observations (X2 and X4 for each source respectively). NGC~4395~ULX-1 and NGC~6946~ULX-1 both have a single brighter observation which exceeds $L_{X}=1\times10^{39}$\,erg\,s$^{-1}$ (C1 and X3 for each source respectively).

\begin{figure}[!h]
\begin{center}
\vspace{-1cm}
\includegraphics[width=9cm]{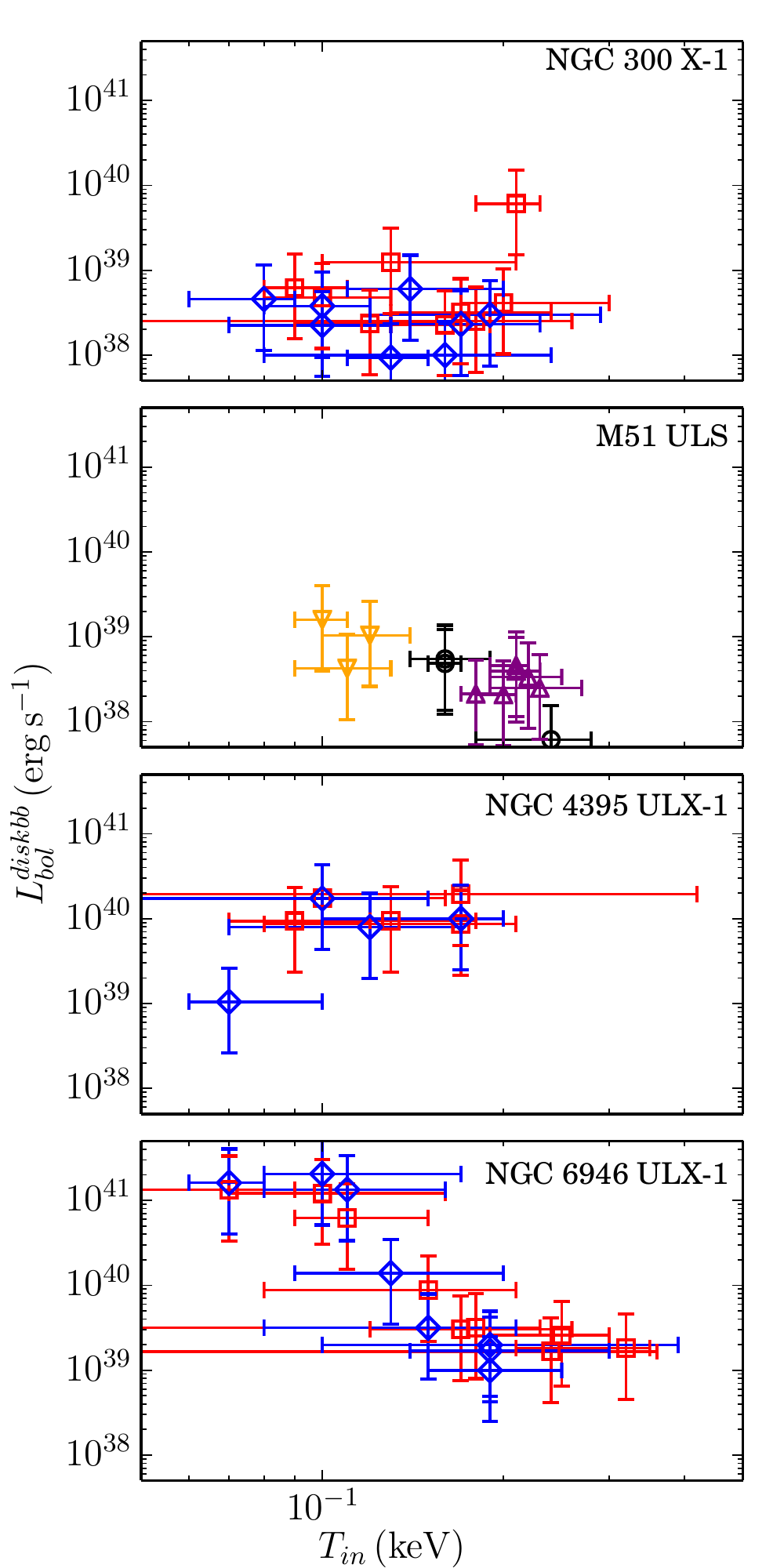}
\end{center}
\vspace{-0.3cm}
\caption{The inner disc temperature and bolometric luminosity of the fitted accretion disc component of the best-fitting MCD models for M51~ULS, and the best-fitting physically-motivated models for NGC~300~X-1, NGC~4395~ULX-1 and NGC~6946~ULX-1. Black circles indicate a simple MCD model, yellow downward-pointing triangles indicate a MCD model with an additional {\tt mekal} component, purple upward-pointing triangles indicate a MCD model with an {\tt edge} component, red squares indicate a {\tt diskbb*simpl} model with suitable residual components, and blue diamonds indicate a {\tt diskbb+comptt} model with suitable components.}
\vspace{-0.3cm}
\label{fig:templum}
\end{figure}

We show the luminosity-temperature relation for the {\tt diskbb} component across the various physically-motivated models for our sources in Fig.~\ref{fig:templum}. The bolometric luminosity was calculated from the model normalisation of the {\tt diskbb} fit. For M51~ULS, we make a distinction between those models which used a {\tt mekal} component to fit the residuals, those that used an {\tt edge} component, and those that were not improved with an additional residual component. None of the groups show any evidence of a trend except for the three observations fitted only with a MCD, which shows a very tentative inverse relation between the disc luminosity and temperature. For the other sources, we distinguish between the models that use a {\tt simpl} component and those that use a {\tt comptt} component. Neither NGC~300~X-1 nor NGC~4395~ULX-1 show a strong relation, however NGC~6946~ULX-1 shows a clear, statistically significant ($>3\sigma$) inverse relationship between disc luminosity and temperature for both types of model. Linear regression analysis shows the relations to be $L_{\rm bol}\propto T_{\rm in}^{-3.5\pm0.5}$ for {\tt simpl} models and $L_{\rm bol}\propto T_{\rm in}^{-5.8\pm0.9}$ for {\tt comptt} models. (A similar inverse relationship exists if the spectra are instead fitted with a {\tt bbody} model, with more-or-less consistent slopes).

\begin{figure}
\begin{center}
\vspace{-0.1cm}
\includegraphics[width=9cm]{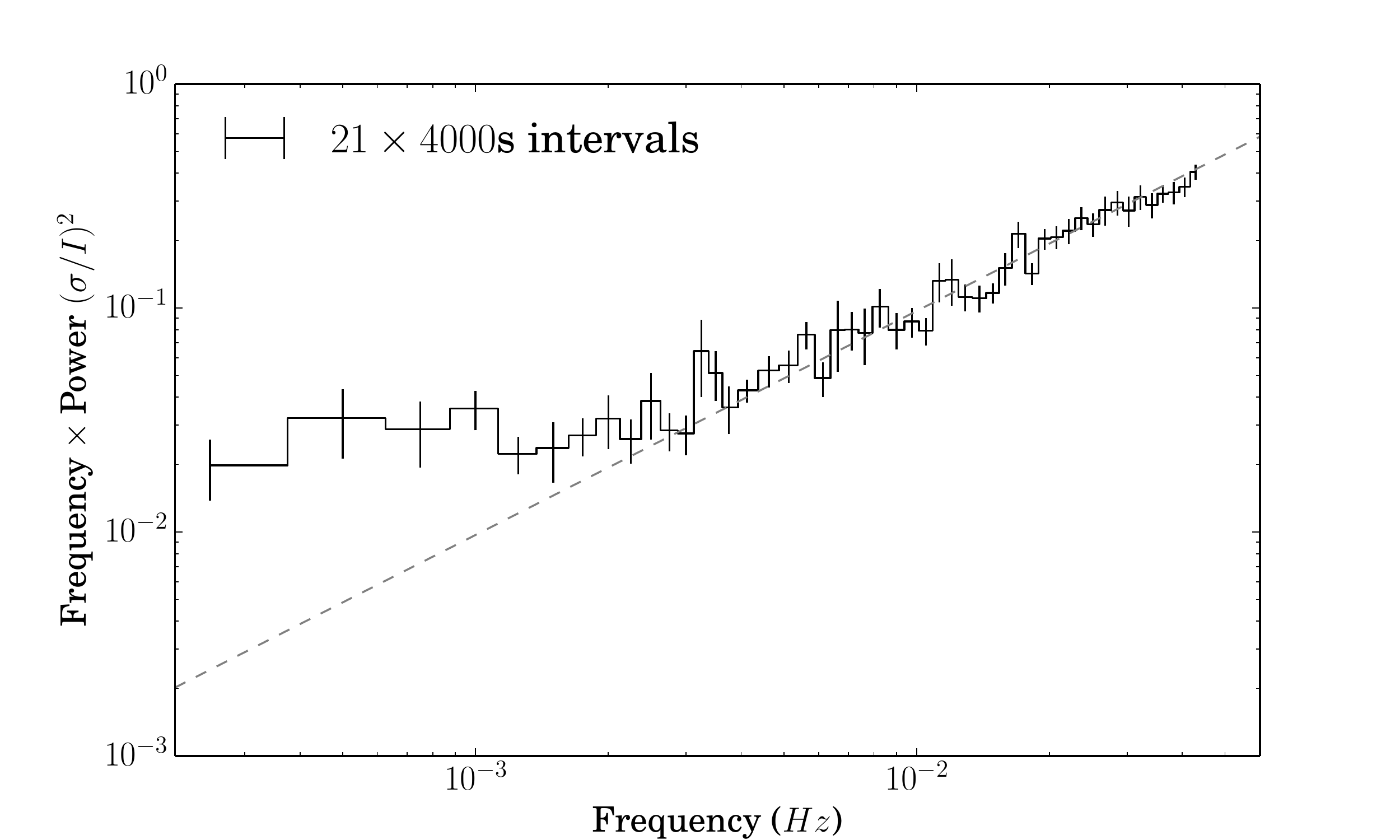}
\vspace{-0.1cm}
\includegraphics[width=9cm]{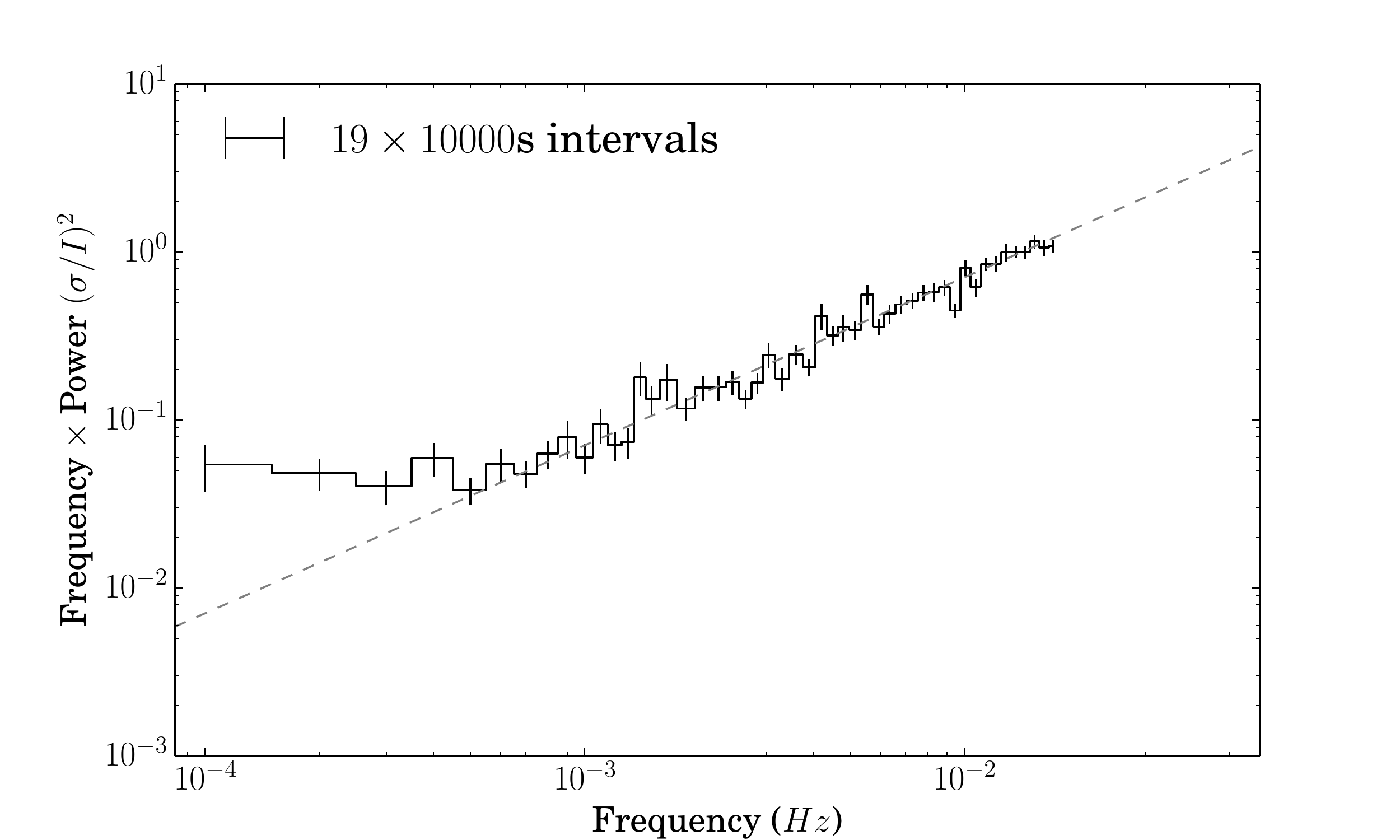}
\end{center}
\vspace{-0.2cm}
\caption{The \xmm~(top) and \chandra~(bottom) power spectra of NGC~300~X-1, normalised so that the power is in units of the squared fractional rms per frequency interval. The grey dashed line indicates the white noise level, and error bars are the standard error on the mean. The spectrum is geometrically rebinned with a rebinning co-efficient of 1.05.}
\vspace{-0.3cm}
\label{fig:powspec}
\end{figure}

\begin{figure}
\begin{center}
\vspace{-0.1cm}
\includegraphics[width=9cm]{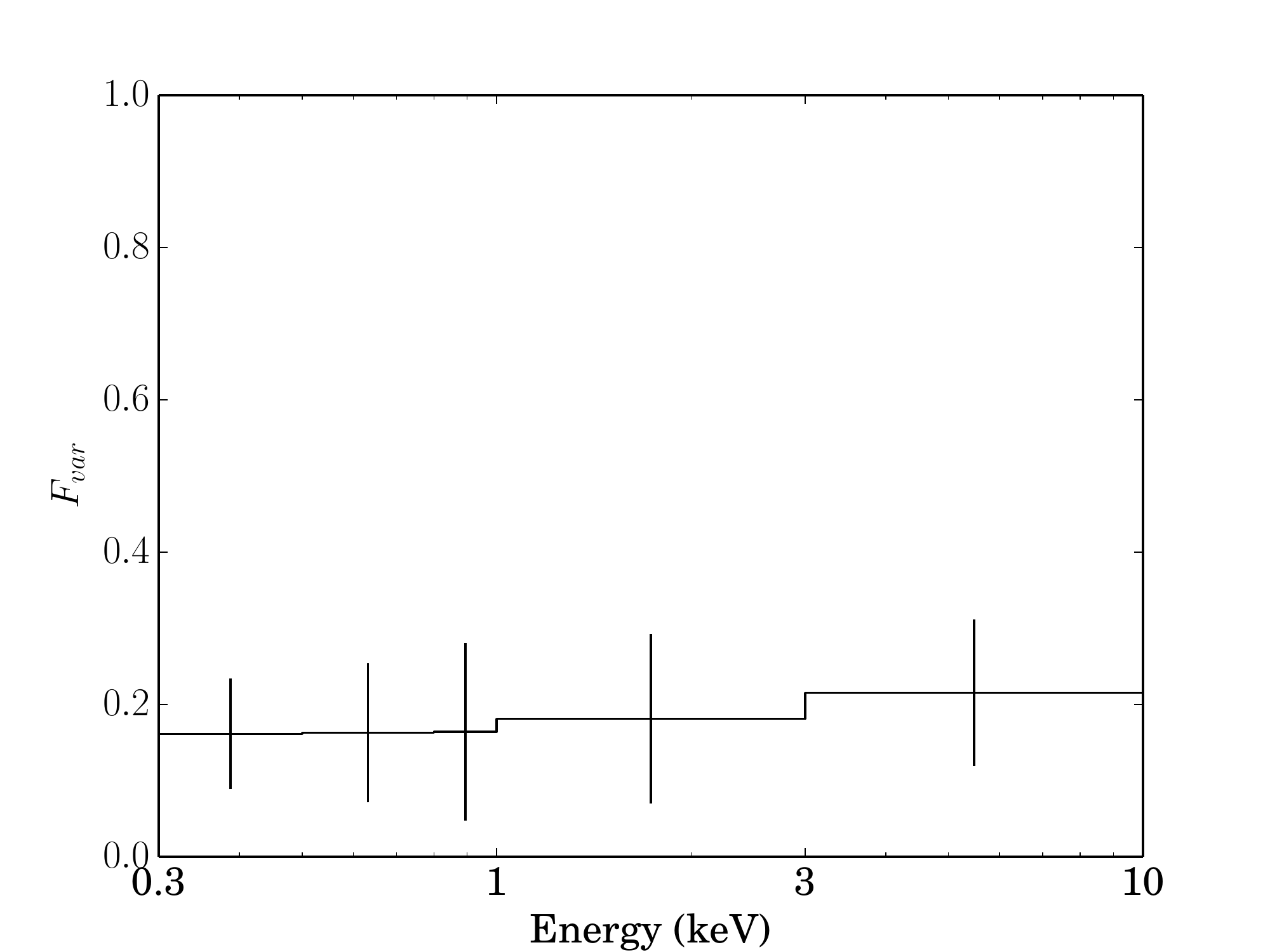}
\vspace{-0.1cm}
\end{center}
\caption{The \xmm~fractional rms variability spectrum for NGC~300~X-1, using the energy bands 0.3--0.5, 0.5--0.8, 0.8--1.0, 1.0--3.0 and 3.0--10.0\,keV. The errors are calculated by propagating the standard error on the mean for each power spectrum bin summed to find the fractional variability.}
\vspace{-0.3cm}
\label{fig:rmsspec}
\end{figure}

The power spectra for NGC~4395~ULX-1 and NGC~6946~ULX-1 across all of their observations are consistent with the Poisson noise level down to $\sim10^{-4}$\,Hz. At the longest timescales, $<10^{-4}$\,Hz, we see some evidence for red noise in the \chandra~power spectrum of M51~ULS. NGC~300~X-1 is the only source where there is sufficient signal to see a clear red noise component in both the \xmm~and \chandra~power spectra (Fig.~\ref{fig:powspec}) with slope $\alpha\sim1$ for $P(\nu) \propto \nu^{-\alpha}$, consistent with the findings by \citet{barnard08} and characteristic of an accretion disc. 

Since the timing data quality for M51~ULS, NGC~4395~ULX-1 and NGC~6946~ULX-1 is poor, we are not able to place good constraints on the fractional rms for most observations. We obtain upper limits on the fractional rms of $\lesssim0.35$ for M51~ULS, with \xmm~observations X2 and X4 having $0.39\pm0.06$ and $0.33\pm0.04$ respectively, consistent with most observations having a high variability of $\sim30\%$. Likewise, most observations of NGC~4395~ULX-1 and NGC~6946~ULX-1 give upper limits of $<0.3$, with observation X3 for NGC~4395~ULX-1 having fractional rms $0.09\pm0.01$ and observation X4 for NGC~6946~ULX-1 having an upper limit to the fractional rms of $<0.15$. This could be consistent either with the sources being moderately variable, or having little to no variability at all.

Since NGC~300~X-1 has sufficient variability power for further analysis, we created a fractional rms spectrum to examine the variability across different energy bands (Fig.~\ref{fig:rmsspec}). We did this by integrating over the \xmm~power spectrum in five different energy bands. Across the entire energy range we examine, the fractional variability remains constant at $\sim0.2$, showing no evidence of strong energy-dependence.

\subsection{Optical Counterparts}
\label{sec:opt}

Using our own astrometric corrections and error circles, we successfully identify the same stellar counterparts for our sample that previous studies do, except for NGC~6946~ULX-1 for which we present the first identified counterpart. In all cases we use STMAG units.

For NGC~300~X-1, we identify the Wolf-Rayet star but not the other stars suggested in \citet{binder15} as potential counterparts. We see one other bright possible counterpart, but since we also find the Wolf-Rayet star, it is likely that it is the genuine companion star. We find a single counterpart to NGC~4395~ULX-1, the same as identified in \citet{vinokurov16}. For M51~ULS, we comfortably identify the counterpart found in \citet{terashima06} using both methods of astrometry matching, therefore we conclude that this is most likely to be the genuine counterpart. We do however note that the error circle obtained from USNO-matching also contains a number of other bright blue stars, as well as probable red giants or AGB stars, and the error circle obtained from direct {\it Chandra} and {\it HST} matching contains further blue sources and a probable star cluster, so there do exist a number of other legitimate ULX counterparts for this source.

For NGC~6946~ULX-1, we identify a single star as a possible counterpart within the error circle, which is only detected in the {\textit F814W} band, with magnitude $m_{F814W} = 22.3\pm0.7$ after correcting for extinction, which corresponds to an absolute magnitude of $M_{F814W}=-6.4$ using the distance to NGC~6946 given in Table~\ref{tab:sources}. Given that we only have access to data in the $I$ band and the narrow H$\alpha$ band, we acknowledge that it is possible that a very blue optical counterpart may also be present within the error circle but undetected. Shorter-wavelength optical and UV bands would be required to further characterise potential companion stars to NGC~6946~ULX-1.

\section{Discussion}
\label{sec:disc}

We have identified a heterogenous sample of four soft, bright sources observed with luminosities just below the Eddington luminosity for a $\sim$10\,$M_{\odot}$ BH. Since we are only looking at the sources with the highest-quality data and selection effects bias us towards soft sources, these objects are not representative of this luminosity range as a whole. Despite this they are still of interest, and even within this small sample we observe different accretion behaviours, including the highest-luminosity canonical states and potential ultraluminous supersoft sources.

\subsection{NGC 300 X-1: the highest luminosity canonical states}
\label{sec:vhs}

The soft power-law dominated spectrum of NGC~300~X-1 with $\Gamma\sim2.5$ and a fairly high fractional rms of $\sim0.2$ are typical features of the canonical very high/steep PL accretion state of stellar mass BHs. The observation that appears to be an exception to this interpretation is X2, whose lower luminosity and harder PL slope when fitted with a MCD and PL model potentially indicate the source dropping into a hard state. 

The broad Gaussian line at 0.9\,keV offers a significant improvement in the spectral fit in the case of three of the \xmm~observations. \citet{carpano07} suggest that these could be unresolved emission lines, possibly from reprocessing by a photoionised stellar wind -- they are not present in all observations, but tend to be present in the higher-luminosity ones. \citet{binder15} find similar improvements in the fit to ours using an {\tt apec} thermal plasma model. This feature is consistent with being a combination of emission and absorption lines produced by an extended hot gas corona.

As NGC~300~X-1 is already a well-studied source, we do not go into much further discussion of its properties. \citet{binder15} suggest that while spectral models find low inner disc temperatures, since the mass of the BH is known and this would thus imply a strong retrograde BH spin, it is more likely that there is a hot inner disc covered by an extended corona that is able to access cooler seed photons from further out in the accretion disc. Fitted accretion discs appearing cooler and less luminous than expected due to accretion energy being emitted through the corona is a normal feature of the steep PL state observed in other sources (e.g. \citealt{kubota16}), as well as an amount of disc truncation compared to the standard thermal-dominant state. This may also be why there does not seem to be a clear accretion disc luminosity-temperature relationship.

While in Galactic sources the steep PL state is often considered to be a transitional state between the low/hard state and the thermal dominant state, it is possible that in NGC~300~X-1 the high mass of the companion star means that the stellar wind can provide it with a persistent rapid rate of fuel supply and therefore enable it to possess a stable accretion rate, at $\sim10\%$ of the Eddington luminosity. 

\subsection{A selection of highly luminous supersoft sources}
\label{sec:uls}

The remaining three sources all show behaviour distinct from the canonical sub-Eddington accretion states. NGC~4395~ULX-1 and NGC~6946~ULX-1 both show a very steep power-law spectrum, with $\Gamma>3$ even when considering the presence of an accretion disc and accounting for residuals. M51~ULS is consistent with a single thermal component spectrum, and does not feature the steep hard tail found in the other sources. 

The first obvious comparison that can be drawn is with ultraluminous supersoft sources (ULSs), such as M101~ULX-1 \citep{soria16} and NGC~247~ULS \citep{feng16}. Characterised by super-Eddington bolometric luminosities and dominated by a low temperature thermal component with $T\sim0.1$\,keV, ULSs may at first glance seem to be evidence of IMBHs in a thermal-dominated accretion state. However, the luminosity, temperature and expected accretion rate for the thermal-dominated accretion state are not consistent with an IMBH interpretation. Instead, \citet{soria16} suggest that these sources are examples of the supercritical ULX model \citep{poutanen07} viewed through very optically thick outflowing winds, at very extreme accretion rates and/or viewed at a high inclination, such that all hard photons from the central source are downscattered in the wind, removing any hard tail in the spectrum. This means that the blackbody shape in the spectrum does not actually originate from the disc but from the wind -- with very low peak temperatures, a MCD model and a standard blackbody model are statistically indistinguishable in the {\it XMM-Newton} data, so this is acceptable in the context of our spectral fitting.

We note here that in the case of an interpretation as a super-Eddington accreting compact object, the object's underlying nature could be either a BH or a neutron star (as a small number of ULXs have been discovered to be pulsars e.g. \citealt{bachetti14, fuerst16, israel16}). The detection of pulsations is to date the only definitive method of confirming the presence of a neutron star rather than a BH, and since we do not detect them in any of our sources we continue our discussion under the assumption of super-Eddington accretion onto a BH while conceding that there is a possibility that these sources are neutron stars -- a possibility we are unable to test at this time.

For M51~ULS, it is not likely that we are viewing the accretion disc of an IMBH, since the luminosity is too low to be in the sub-Eddington thermal-dominant accretion state for the disc temperature we observe. Additionally, while no strong trend is observed, there is tentative evidence of an inverse relationship between the disc component's luminosity and temperature, as opposed to the $L_{\rm disc}\propto T_{\rm in}^4$ expected from an accretion disc in the thermal dominant state, and similar to inverse relationships seen in other ULSs (e.g. \citealt{urquhart16}) and in the soft component of some ULXs (e.g. \citealt{feng07,soria07,kajava09}). The spectrum is consistent with viewing a supercritically accreting source through an optically thick wind that has downscattered the seed photons into a thermal spectrum. While it is hard to constrain the source variability, the fractional rms upper limit for M51~ULS is generally around 0.3, consistent with the strong short-term variability seen in other ULSs. It exhibits an absorption edge at $\sim1$\,keV in some of our observations of the source, and we note that transient absorption edges are a feature seen in a number of ULSs as well (e.g. \citealt{feng16, urquhart16}) and are further evidence of absorption in an effectively optically thick outflow wind. All these features are consistent with M51~ULS being a normal member of the ULS population -- and indeed it makes an appearance as one of seven ULSs examined in \citet{urquhart16}.

What may be more challenging to explain are the PL spectral shapes of NGC~4395~ULX-1 and NGC~6946~ULX-1. For most ULSs that sometimes exhibit hard residuals (e.g. M101~ULX-1), these are transient -- in the case of M101~ULX-1, \citet{soria16} fit them with multiple {\tt mekal} components and suggest that they may be the result of central source emission being observed through gaps in the clumpy wind that makes up the outflow. However, the steep slopes of our two sources are persistent and cannot simply be fitted with additional {\tt mekal} components, instead requiring a PL shaped model. 

Perhaps more similar is NGC~247~ULS, which is observed to have a steep PL slope with $\Gamma=3.9\pm0.4$ in its 2014 observation. \citet{feng16} suggest that this may be an extreme example of the soft ultraluminous regime, with the hard emission very suppressed so that it manifests as a steep PL and no spectral turnover is evident (either because it is not present, or else because the hard flux is far too low for one to be detected). They also observe moderate short-term variability, which appears to originate from the soft component itself, rather than variability observed primarily in the hard emission as the hard central emission is occulted by a clumpy wind, as in the soft ultraluminous regime \citep{middleton15a}.

While there is no strong relationship between the bolometric luminosity and the temperature of the disc component for NGC~4395~ULX-1, there is an inverse relationship between those properties for NGC~6946~ULX-1, which is inconsistent with being produced by an accretion disc. It could, however, be produced instead by an expanding and contracting photosphere, and bears similarity with 'standard' ULS behaviour and that of the soft excess in ULXs, which would lend support to these very steep PL sources being intermediate objects between a soft ultraluminous ULX and a ULS.

If this is the case, the question remains of why M51~ULS and other typical ULSs have such high levels of variability compared to this moderately variable steep PL ultraluminous state we see in NGC~4395~ULX-1 and NGC~6946~ULX-1 if little to no hard emission emerges at all. \citet{middleton15a} predicts that at very high inclinations, the density of the clumps in the wind would smooth out any variability caused by obscuration of the central source. It is possible then that at extreme accretion rates, while the hard central emission is completely obscured, the outflowing wind varies in temperature by radius such that the clumpy edge of the wind still imprints variability upon the spectrum by occulting deeper parts of the wind at different temperatures to the clumps. Alternatively, \citet{feng16} suggest that at very high inclinations, the wind or photosphere itself may be masked by an uneven occulter at much larger radii, such as the outer edges of a warped accretion disc or even a circumbinary disc.

It is appealing to explain NGC~4395~ULX-1 and NGC~6946~ULX-1 as a special case of the soft ultraluminous regime, in which the hard central photons are mostly but not completely downscattered, leaving a steep PL tail and therefore placing them within an ultraluminous unified model. However, the steepness of the PL tail of their spectra and the lack of a spectral turnover brings into question whether the explanation is that simple. These two ULSs, as well as the similar observation of NGC~247~ULS, exhibit PL slopes with $\Gamma=3$--4, whereas the majority of soft ultraluminous sources have a slope of $\Gamma=2$--3 when fitted either with a single-component PL or a MCD+PL model \citep{gladstone09}. Even an exception found by \citet{gladstone09} to have $\Gamma>3$ (NGC~5408~X-1) very clearly exhibits a high-energy turnover as well as a steep PL slope.

{\it NuSTAR} data of ULXs has shown that at energies above the hard spectral turnover, the spectrum is consistent with a steep PL, with $\Gamma\sim3$ (e.g. \citealt{walton14,walton15,mukherjee15}). Therefore it is possible that instead of being absent in NGC~4395~ULX-1 and NGC~6946~ULX-1, this hard turnover is present at an unusually low energy ($\sim1$\,keV) and is therefore masked by the soft component and possibly also residual features present in the spectrum. The issue is then why the turnover is at such a low energy, instead of why the PL is so steep. In order to produce such a spectrum using a {\tt comptt} model, the plasma temperature is required to be very cool with $kT<1$\,keV (below the general applicability of the {\tt comptt} model). Rather than as actual Comptonisation, this component is often interpreted as a distorted hot accretion disc -- however, in this case, this would imply a observed disc temperature not much higher than the temperature of the outflowing wind, which is not unexpected for a system in which we do not see the hot inner regions of the disc. In any case, we do not have the statistics to test either of these scenarios at the present time, so we also examine a number of other possible interpretations of this steep, unbroken PL tail for completeness.

It could be possible that these steep PL tails are direct emission from the centre of the accreting system and are the result of Compton upscattering in addition to a downscattering outflowing wind, in which case the {\tt comptt} model fits indicate that the Comptonising medium is optically thick and in most cases low temperature. However, it is unclear how this medium, if similar to the accretion disc corona observed in the steep PL state in sub-Eddington sources, could possibly be seen if the source is at a high inclination or otherwise dominated by the outflowing wind. Additionally, it would require a source of low-temperature photons to be upscattered in the first place -- we would expect the central source to be high-energy, with the low-energy photons originating in the outflowing wind, outside the Comptonising medium.

\begin{table*}
\begin{minipage}{172mm}
\caption{The properties of the most likely optical counterparts of our sources.} \label{tab:counterparts}
\begin{center}
\begin{tabular}{lcccccccl}
  \hline
  Name & Optical R.A. \& Dec. & $m_{435}^a$ & $m_{555}^a$ & $m_{606}^a$ & $m_{814}^a$ & $M_V^b$ & $M_I^c$ & Reference \\
   & (J2000) & & & & & & & \\
  \hline
  NGC~300~X-1 & 00 55 09.99 $-37$ 42 12.65 & ... & ... & 22.41 & 22.33 & $-4.2$ & $-4.0$ & \citet{binder15} \\
  M51~ULS & 13 29 43.31 $+47$ 11 34.73 & 23.20 & 24.01 & ... & 25.50 & $-5.5$ & $-4.2$ & \citet{terashima06} \\
  NGC~6946~ULX-1 & 20 35 00.42 $+60$ 09 07.1 & ... & ... & ... & 22.3 & ... & $-6.2$ & This work \\ 
  \hline
   &  & $m_{275}^a$ & $m_{336}^a$ & $m_{438}^a$ & $m_{547}^a$ &  &  &  \\
  \hline
  NGC~4395~ULX-1 & 12 26 01.44 $+33$ 31 31.1 & 19.97 & 20.50 & 22.08 & 22.26 & $-6.2$ & ... & \citet{vinokurov16} \\
  \hline
\end{tabular}
\end{center}
$^a$Extinction-corrected stellar magnitudes of the brightest optical counterpart in STMAG.\\
$^b$Absolute magnitude in the $V$ band, either calculated from the photometry (M51~ULS, NGC~4395~ULX-1) or estimated from spectral type (NGC~300~X-1).\\
$^c$Absolute magnitude in the $I$ band calculated from the photometry.
\end{minipage}
\end{table*}

Alternatively, as it expands, the outflowing wind could eventually disperse and become optically thin. If the material beyond the photosphere is still very hot and ionised, it may be sufficient for Compton upscattering of the soft emission from the outer regions of the accretion disc or wind photons from its optically thick phase. There is some degeneracy in the parameters of the {\tt comptt} model in that a similarly steep PL slope can be produced with an optically thin, high-temperature plasma, although model fits tend to prefer optically thick, low-temperature parameter values. However, this would require the wind to be at a very high temperature, inconsistent with current measurements.

Another possibility is that the hard emission is caused by shocks within the expanding envelope of the outflow between shells and/or clumps of different densities produced as the wind varies, or a collision between the outflow and a dense interstellar medium generated by the stellar wind from the companion star. Collisional shocks have been suggested as a mechanism in ULXs before, in the context of explaining the origin of soft emission line residuals within the spectra (e.g. \citealt{pinto16}), however these shocks appear to produce a thermal, emission-line rich spectrum and not a non-thermal PL-shaped spectrum.

It may be illuminating to consider soft Galactic objects with similarly steep PL spectra for insight into the underlying physics of our sources. We can draw parallels with Galactic microquasars that enter a hypersoft state, such as Cyg~X-3 and GRO~J1655-40, characterised by very steep PL slopes with $\Gamma=4$--8. Study of Cyg~X-3's hypersoft state revealed that while there was no hard X-ray emission detected, the source did emit $\gamma$-rays while in that state, indicating that highly-energetic processes were still taking place albeit obscured in the hard X-rays \citep{koljonen10}. SED modelling of GRO~J1655-40 suggests that its unusually steep spectrum is caused by a powerful Compton-thick and ionised disc wind, driven by a source accreting at a near- or super-Eddington intrinsic luminosity \citep{uttley15, shidatsu16}. This interpretation ties the hypersoft state into the ULX family in a similar fashion to ULSs, and demonstrates that such sources can manifest with steep PL slopes, just as NGC~4395~ULX-1 and NGC~6946~ULX-1 do. Additionally, the absorption edges observed in the spectra of NGC~6946~ULX-1 indicate high levels of ionisation.

The main difference between the hypersoft accretion state and our very steep PL objects is the transience of the hypersoft state, compared with our sources which are for the most part persistent in their steepness. This could be attributed to their companion stars allowing persistent accretion through Roche lobe overflow, keeping the source in a single accretion state across years of observation. Additionally, a direct comparison of these sources with GRO~J1655-40 shows the Galactic source to have a much higher blackbody temperature of $\sim0.4$\,keV, making drawing direct parallels between the objects tricky.

We must also consider that NGC~4395~ULX-1 and NGC~6946~ULX-1 may be similar to NGC~300~X-1, in a sub-Eddington steep PL state, albeit with a steeper PL than is usually expected in such a source, implying a particularly low plasma temperature in the Comptonising corona. They are consistent with having moderately high variability and a cool disc temperature similar to NGC~300~X-1. However, their luminosities are for the most part 3--5 times higher that that observed for NGC~300~X-1 while in the steep PL state, therefore we must conclude that for this analogy to hold, the sources either have a higher BH mass (60--100\,$M_{\rm BH}$) or are consistently accreting at a higher Eddington fraction than NGC~300~X-1 (30--50\%).

In Table~\ref{tab:counterparts} we collate previous studies of the optical counterparts of our sources with this work. All sources have very bright counterparts, similar to ULX counterparts, whose absolute magnitudes tend to lie in the range of $-4 > M_V > -8$, with most having $M_V \approx -6$ \citep{vinokurov16}. Those counterparts for which we have a value of $M_V$ lie in this range and have possible supergiant spectral types, with the magnitudes of M51~ULS and NGC~300~X-1 being consistent with an OB supergiant and WR star respectively. NGC~4395~ULX-1 has an optical spectrum with similar features to other ULX spectra \citep{vinokurov16}, which are very blue and show features such as broadened He~{\sc ii}, H$\alpha$ and H$\beta$ emission lines. While these spectra are similar to late-nitrogen WR star or OB supergiant spectra, \citet{fabrika15} suggest that these may actually be dominated by emission from an irradiated supercritical accretion disc wind, similar to Galactic source SS~433. If this is the case, this would be further support for NGC~4395~ULX-1 being a super-Eddington accreting object. For the counterpart to M51~ULS, at a less extreme magnitude, it is more ambiguous as to whether the counterpart is dominated by a supergiant star or a supercritical disc -- optical spectroscopy would be required to make this distinction, and to determine whether the spectrum is suitable for dynamical mass measurements.  

The $I$-band magnitude of the counterpart of NGC~6946~ULX-1 is much brighter than the other counterparts we observe. This could possibly be an indiciation of a red supergiant companion star, found in some ULX systems (e.g. \citealt{heida14, heida16}). Red supergiant companion stars are useful for radial velocity measurements as they are expected to be dominant over the accretion disc in the near infrared, since the contribution from the accretion disc is lower in that regime than in the optical. Therefore if this counterpart is indeed a red supergiant, it could prove a good target for spectroscopic mass measurements.

\section{Conclusions}
\label{sec:conc}

We have searched a new, clean catalogue of extragalactic, non-nuclear X-ray point sources for luminous objects with peak luminosity below $3\times10^{39}$\,erg\,s$^{-1}$ in order to find sources at the Eddington Threshold and to probe the nature of accretion at or just below super-Eddington rates. We identified a heterogenous sample of four sources, all very soft due to the nature of our selection methods:

$\bullet$ NGC~300~X-1 is a well-known example of a BH accreting in a persistent steep PL state, made possible by its high-mass WR companion star. Its selection does not come as a surprise given that we would expect the highest-luminosity examples of the canonical sub-Eddington BH accretion states to appear at Eddington threshold luminosities. 

$\bullet$ M51~ULS is best fitted with a MCD disc model, with a blackbody-shaped spectrum. This spectral shape and its likely fairly high levels of fractional variability make it similar to ULSs, so this may be a super-Eddington accreting source viewed at high inclinations and/or through an optically thick outflowing wind, which causes both the lack of high-energy photons, which are beamed away or downscattered, and the lower than expected luminosity.

$\bullet$ NGC~4395~ULX-1 and NGC~6946~ULX-1 both exhibit a very steep PL tail, a complication that does not make them immediately comparable with ULSs. We consider a number of possible physical scenarios, including Compton-upscattering at the point the wind becomes optically thin, shocks within the wind or where the wind meets the interstellar medium, or manifestations of the steep PL state with relatively low corona temperatures. However, we conclude that it is most likely that they are an intermediate stage between ULSs and the soft ultraluminous regime in ULXs, in which most but not all of the hard central emission is downscattered, leaved a very steep PL tail. They also bear similarities with the hypersoft state observed in a handful of Galactic microquasars, which may also be associated with super-Eddington accretion driving a Compton-thick wind.

Even with the best-quality data sets, many of these sources are not luminous enough to provide sufficient data for highly in-depth investigation unless they are targeted by long dedicated observation campaigns. Additionally, due to selection bias, our sample is not representative of the Eddington Threshold population as a whole. However we have demonstrated even with this limited sample that the Eddington Threshold population exhibits a variety of accretion behaviours that are not well understood, and would be an excellent regime for investigation by future X-ray missions such as {\it Athena}.

\section*{Acknowledgements}

We would like to thank our referee, Roberto Soria, for useful comments on this paper. We gratefully acknowledge support from the Science and Technology Facilities Council (HE through studentship grant ST/K501979/1 and TR as part of consolidated grant ST/L00075X/1).

The scientific results reported in this paper are based on data obtained from the {\it Chandra} Data Archive, and on archival observations obtained with {\it XMM-Newton}, an ESA science mission with instruments and contributions directly funded by ESA Member States and NASA.

\bibliography{ssulxpaper}
\bibliographystyle{mn2e}
\bsp

\label{lastpage}

\end{document}